# The Origin and Evolution of Saturn, with Exoplanet Perspective

Sushil K. Atreya, Aurélien Crida, Tristan Guillot, Jonathan I. Lunine, Nikku Madhusudhan, and Olivier Mousis

**Abstract**

Saturn formed beyond the snow line in the primordial solar nebula that made it possible for it to accrete a large mass. Disk instability and core accretion models have been proposed for Saturn's formation, but core accretion is favored on the basis of its volatile abundances, internal structure, hydrodynamic models, chemical characteristics of protoplanetary disk, etc. The observed frequency, properties and models of exoplanets provide additional supporting evidence for core accretion. The heavy elements with mass greater than $^4$He make up the core of Saturn, but are presently poorly constrained, except for carbon. The C/H ratio is super-solar, and twice that in Jupiter. The enrichment of carbon and other heavy elements in Saturn and Jupiter requires special delivery mechanisms for volatiles to these planets. In this chapter we will review our current understanding of the origin and evolution of Saturn and its atmosphere, using a multi-faceted approach that combines diverse sets of observations on volatile composition and abundances, relevant properties of the moons and rings, comparison with the other gas giant planet, Jupiter, analogies to the extrasolar giant planets, as well as pertinent theoretical models.

## 2.1 Introduction

Saturn, though about one-third the mass of Jupiter, is the largest planetary system in the solar system, considering the vast reach of its rings and dozens of known moons. Thus, Saturn is key to understanding the origin and evolution of the solar system itself. Models, observations, comparison with Jupiter, the other gas giant planet, and analogies with extrasolar giant planets have begun to give a sense of how Saturn, in particular, and the giant planets in general, originated and evolved.

Two distinct mechanisms of giant planet formation have been proposed in the literature: (1) disk instability (or "grey" instability) and (2) nucleated instability (or core accretion). The latter goes back to papers by Hayashi (1981) and his colleagues (e.g. Mizuno, 1980), and requires the accretion of a solid body (rock/metal, ice, and possibly, refractory organics) up to a critical mass threshold at which rapid accretion of gas becomes inevitable—typically 10 times the mass of the Earth (see Armitage, 2010, for a discussion). The former had its origin in the 1970's (see Cameron, 1979) for hot, massive disks, but it was determined later (Boss, 2000; Mayer et al., 2002) that the instabilities required to break up a portion of a gaseous disk into clumps are a feature of cold, massive disks. We focus on each of these contrasting models in turn, and then discuss the observational indicators in our own and extrasolar planetary systems that might distinguish between the two models.

The disk instability model is based on numerical simulations showing that massive, relatively cold disks will spontaneously fragment due to a gravitational instability, leading to multiple discrete, self-gravitating masses. In computer simulations of the process these features seem somewhat ill-defined, and it is not possible to track the subsequent condensation of these features in the same hydrodynamical simulation that tracks the onset of the instability itself. Nonetheless, basic disk physics dictates that such fragmentation will occur for a sufficiently massive or cold disk (Armitage, 2010), and that the timescale for the fragmentation once the instability does occur is extremely short—hundreds to thousands of years.

Once formed, the fragments (assuming they continue to contract to form giant planets) are usually sufficiently numerous that the aggregate planetary system is dynamically unstable. The planets will gravitationally interact, scattering some out of the system and leaving the others in a variety of possible orbits. The evidence from microlensing of a substantial population of free-floating Jupiter-mass objects (Sumi et al., 2011) − not associated with a parent star − constitutes one argument in favor of the importance of this formation mechanism.

On the other hand, it is not evident how giant planets formed by the disk instability mechanism acquire significant amounts of heavy elements over and above their parent star's abundances. It has been argued that subsequent accretion of planetesimals would generate the increased metallicity, but the disruption of the disk associated with the gravitational instability might have removed the raw material for large amounts of planetesimals—the materials going into numerous giant planets that are then kicked out of the system. A subsequent phase of disk building or direct accretion of planetesimals from the surrounding molecular cloud may have to be invoked. And this begs the question of core formation − giant planets formed in this way may have super-solar metallicities but lack a heavy element core unless (as seems unlikely) very large (Earth-sized) planets are consumed by these objects.

The core accretion model, in contrast, begins by building a heavy element core through planetesimal and embryo accretion in the gaseous disk (embryo is usually reserved for lunar-sized bodies and upward). At some point, the gravitational attraction of the large core leads to an enhanced accretion of gas, so much so that gas accretion quickly dominates in a runaway process and the object gains largely nebular-composition gas until its mass is large enough to create a gap in the disk and slow accretion. Such a model produces, by definition, a heavy element core, and through co-accretion of gas and planetesimals an envelope enrichment of heavy elements as well. The model's Achilles heel is the time required to build the heavy element core to the point where rapid gas accretion occurs—millions of years or more. The onset of rapid gaseous accretion, by which point further growth may be rapid, depends not only on the core accretion rate but also, through the critical core mass (roughly 10 $M_E$, where $M_E$ is an Earth mass) needed to trigger rapid gas accretion, the envelope opacity and hence metallicity. Furthermore, the core accretion rate itself is a sensitive function of what one assumes about the planetesimal size distribution and surface density in the disk.

A plausible timescale for the formation of Saturn must be consistent with the lifetime of gas in disks, but may also be constrained by the 3-5 million year estimate of the formation duration of Iapetus from its geophysical shape and thermal history (Castillo-Rogez et al., 2009). Earliest models had lengthy formation times (e.g. 8 Myr; Pollack et al., 1996) but more recent models can make Saturn in a few million years by appropriate selection of nebular parameters such as grain distribution and opacity (Dodson-Robinson et al., 2008).

The overall history of the solar system and presence of a substantial terrestrial planet system inward of Jupiter and Saturn suggests that the extreme dynamical scattering suffered after disk instability protoplanets are formed did not happen in our solar system. Furthermore, if the 3-5 million year estimate of the interval between the formation of the first solids and the formation of Iapetus (Castillo-Rogez et al., 2009) is correct, the disk instability − if it occurred − would have produced Saturn much too soon after (or even before), the first solids in the solar system condensed out. There is sufficient evidence that the first solids, millimeter size chondrules and calcium aluminum inclusions (CAI's) in chondrites, date back to 4.5682 Gyr (Amelin et al., 2010), which provides clear evidence that submicron size interstellar grains were sticking and accumulating to form solids at the very beginning of the solar system.

Measurement by the Juno mission of the water abundance below the meteorology layer in Jupiter, tied to the abundances of other major elements measured by the Galileo probe, will also provide an indication of how much planetesimal material was accreted (Helled and Lunine, 2014), and to some extent, the nature of the carrier species (e.g., Mousis, 2012). Although it is possible to enrich the envelopes of the giant planets even in the disk instability model by adding planetesimals much later, the presence of both a substantial (10 $M_E$) core and envelope enrichment of heavy elements would strongly militate in favor of the core accretion model. Saturn's core mass may be measured by Cassini, but an inventory of the envelope enrichment of heavy elements and measurement of the deep water abundance will have to await a future Saturn probe.

The core accretion model gets a boost also from observational surveys of exoplanets. An analysis of the frequency of planets with different masses, sizes, orbits and host characteristics reveal that a greater percentage of giant planets are found around higher metallicity stars, and smaller planets between Earth and Neptune mass far exceed Jupiter-sized planets (Howard, 2013; Johnson et al., 2010). This is what one would expect if core accretion were prerequisite for planetary formation. Thus, for our planetary system, at least, core accretion seems to make more sense. Trying to constrain detailed formation mechanisms by matching orbital properties is much more difficult because of the profound effects of migration (Mordasini, et al., 2009; Ida et al., 2013 and references therein).

In addition to their occurrence rates and orbital characteristics, the masses, radii, and atmospheric volatile gas compositions of giant exoplanets may also provide important clues regarding their formation processes, and in turn formation of Saturn and Jupiter in the solar system. With rapid advances in spectroscopic observations of exoplanets, a number of gases relevant to formation models, including water vapor, methane and carbon monoxide have been detected in several giant exoplanets (Section 2.5) revealing diversity in chemical abundances. For example, there are some planets (e.g. HD 209458b) with seemingly lower $H_2O$ abundances than expected from solar elemental composition (e.g. Deming et al., 2013, Madhusudhan et al., 2011a, 2014a), while others (e.g. WASP-43b) appear consistent with super-solar $H_2O$ (e.g. Kreidberg et al., 2014). The latter is consistent with super-solar abundance of measured heavy elements in Saturn and Jupiter (Section 2.2.1), with a good likelihood that their original cores were rich in water ice. On the other hand, WASP-12b – which indicates a C/O ratio (≥1) twice solar (~0.5) – argues for a core made up of largely carbon bearing constituents. If this result is confirmed for a multitude of similar exoplanets, it would have important implications for their formation and the formation of the gas giant planets of the solar system. More generally, new theoretical studies are suggesting that the observable O/H, C/H, and hence, C/O, ratios in giant exoplanetary atmospheres can place powerful constraints on their formation and migration mechanisms, as discussed in section 2.5.3.

## 2.2 Observational constraints

The models of Saturn's formation and evolution are constrained by data presently available on the planet's chemical composition and its interior. This section elaborates on each of these aspects, and forms the basis for the discussions in subsequent sections.

### 2.2.1 Elemental composition of Saturn's atmosphere and comparison to Jupiter

The composition of Saturn's atmosphere has been measured by remote sensing from ground-based and earth orbiting telescopes and flyby and orbiting spacecraft for over half a century. These observations have been instrumental in revealing the chemical makeup of Saturn's stratosphere and upper troposphere. As a result, mole fractions of helium (He), methane ($CH_4$) and a number of its photochemical products including methyl radical ($CH_3$), ethane ($C_2H_6$), acetylene ($C_2H_2$), methyl acetylene ($C_3H_4$) and benzene ($C_6H_6$), ammonia ($NH_3$), hydrogen sulfide ($H_2S$), and those species that are in thermochemical disequilibrium in Saturn's upper troposphere and stratosphere such as phosphine ($PH_3$), carbon monoxide (CO), germane ($GeH_4$) and Arsine ($AsH_3$) have been measured to varying degrees of precision. Some of the most precise data have come from observations made by the Cassini spacecraft (Fletcher et al., this book) that attained orbit around Saturn in 2004 and will embark on proximal orbits toward the end of the mission in 2017 (Baines et al., this book).

The abundances of certain heavy elements (m/z >$^4$He) and their isotopes can be derived from their principal chemical reservoirs in the atmosphere. As discussed earlier, heavy elements are key to constraining the models of the formation of Saturn and its atmosphere. Current best data on the abundances of elements relative to hydrogen in Saturn are listed in Table 2.1. As Jupiter, the other gas giant planet in the solar system, is a good analog for Saturn, we list for comparison also the elemental abundances in Jupiter's atmosphere.

Many more heavy elements have been determined at Jupiter in contrast to Saturn because of in situ Galileo Jupiter entry probe measurements in 1995. Enrichment factors of the elements relative to protosolar values are also listed in Table 2.1, using currently available solar elemental abundances from two different sources (Asplund et al., 2009 and Lodders et al., 2009). Further insight into key elemental abundances is given below, and the reader is referred also to the table footnotes.

**Table 2.1.** Elemental abundances in Jupiter and Saturn and ratios to protosolar values

| Elements | Jupiter | Saturn | Sun-Protosolar (Asplund et al., 2009) [a,b] | Jupiter/Protosolar (using Asplund et al., 2009) [a,b] | Saturn/Protosolar (using Asplund et al., 2009) [a,b] | Sun-Protosolar (Lodders et al., 2009) [l] | Jupiter/Protosolar (using Lodders et al., 2009) [l] | Saturn/Protosolar (using Lodders et al., 2009) [l] |
|---|---|---|---|---|---|---|---|---|
| He/H | $7.85\pm0.16\times10^{-2}$ [c] | $5.5–8.0\times10^{-2}$ [h], taken as $6.75\pm1.25\times10^{-2}$ | $9.55\times10^{-2}$ | $0.82\pm0.02$ | $0.71\pm0.13$ (?) | $9.68\times10^{-2}$ | $0.81\pm0.02$ | $0.70\pm0.13$ (?) |
| Ne/H | $1.24\pm0.014\times10^{-5}$ [d] | | $9.33\times10^{-5}$ | $0.13\pm0.001$ | | $1.27\times10^{-4}$ | $0.098\pm0.001$ | |
| Ar/H | $9.10\pm1.80\times10^{-6}$ [d] | | $2.75\times10^{-6}$ | $3.31\pm0.66$ | | $3.57\times10^{-6}$ | $2.55\pm0.50$ | |
| Kr/H | $4.65\pm0.85\times10^{-9}$ [d] | | $1.95\times10^{-9}$ | $2.38\pm0.44$ | | $2.15\times10^{-9}$ | $2.16\pm0.39$ | |
| Xe/H | $4.45\pm0.85\times10^{-10}$ [d] | | $1.91\times10^{-10}$ | $2.34\pm0.45$ | | $2.1\times10^{-10}$ | $2.11\pm0.40$ | |
| C/H | $1.19\pm0.29\times10^{-3}$ [e] | $2.65\pm0.10\times10^{-3}$ [i] | $2.95\times10^{-4}$ | $4.02\pm0.98$ | $8.98\pm0.34$ | $2.77\times10^{-4}$ | $4.29\pm1.05$ | $9.56\pm0.36$ |
| N/H | $3.32\pm1.27\times10^{-4}$ [e]; $4.00\pm0.50\times10^{-4}$ [f] | $0.80–2.85\times10^{-4}$ [j]; $2.27\pm0.57\times10^{-4}$ with $f_{NH_3}=4\pm1\times10^{-4}$ | $7.41\times10^{-5}$ | $4.48\pm1.71$ [e]; $5.40\pm0.68$ [f] | $1.08–3.84$; $3.06\pm0.77$ with $f_{NH_3}=4\pm1\times10^{-4}$ | $8.19\times10^{-5}$ | $4.06\pm1.55$ [e]; $4.89\pm0.62$ [f] | $0.98–3.48$; $2.78\pm0.73$ with $f_{NH_3}=4\pm1\times10^{-4}$ |
| O/H | $2.45\pm0.80\times10^{-4}$ [e] | | $5.37\times10^{-4}$ | $0.46\pm0.15$ (hotspot) | | $6.07\times10^{-4}$ | $0.40\pm0.13$ (hotspot) | |
| S/H | $4.45\pm1.05\times10^{-5}$ [e] | $1.88\times10^{-4}$ [k] | $1.45\times10^{-5}$ | $3.08\pm0.73$ | $13.01$ | $1.56\times10^{-5}$ | $2.85\pm0.67$ | $12.05$ |
| P/H | $1.08\pm0.06\times10^{-6}$ [g] | $3.64\pm0.24\times10^{-6}$ [g] | $2.82\times10^{-7}$ | $3.83\pm0.21$ | $12.91\pm0.85$ | $3.26\times10^{-7}$ | $3.30\pm0.18$ | $11.17\pm0.74$ |

[a] Protosolar values calculated from the solar photospheric values of Asplund et al. (2009, table 1).
[b] According to Asplund et al. (2009), the protosolar metal abundances relative to hydrogen can be obtained from the present day photospheric values (table 1 of Asplund et al., 2009) increased by +0.04 dex, i.e. ~11%, with an uncertainty of ±0.01 dex; the effect of diffusion on He is very slightly larger: +0.05 dex (±0.01). Note that Grevesse et al. (2005, 2007) used the same correction of +0.05 dex for all elements. dex stands for "decimal exponent", so that 1 dex=10.
[c] von Zahn et al. (1998), using helium detector on Galileo Probe; independently confirmed by the Galileo Probe Mass Spectrometer (GPMS, Niemann et al., 1998).
[d] Mahaffy et al. (2000); Kr and Xe represent the sum of all isotopes except for $^{126}$Xe and $^{124}$Xe that could not be measured by the GPMS but are probably negligible as together they make up 0.2% of the total xenon in the sun.
[e] Wong et al. (2004), based on re-calibration of the GPMS data on $CH_4$, $NH_3$, $H_2O$ and $H_2S$ down to 21 bars, using an experiment unit and represents an update of the values reported in Niemann et al. (1998) and Atreya et al. (1999, 2003).
[f] Folkner et al. (1998), by analyzing the attenuation of the Galileo probe-to-orbiter radio communication signal (L-band at 1387 MHz or 21.6 cm) by ammonia in Jupiter's atmosphere.
*Footnotes continued on next page*

After hydrogen, helium is the most abundant element in the universe, the sun and the giant planets. Conventional thinking has been that the current abundance of helium ratioed to hydrogen in the giant planets should be the same as in the primordial solar nebula from which these planets formed and originally the Big Bang in which helium was created. Thus, precise determination of the helium abundance is essential to understand the formation of the giant planets, in particular, and to shed light on the solar nebula and the universe in general. Whereas helium has been measured very accurately at Jupiter by two independent techniques on the Galileo probe (Table 2.1), such is not the case for Saturn. In the absence of an entry probe at Saturn, helium abundance at Saturn was derived from atmospheric mean molecular weight ($\mu$), using a combination of the Voyager infrared spectrometer (IRIS) and the radio science (RSS) investigations. RSS measured radio refractivity that provides the information on $T/\mu$, where T is the temperature measured by both instruments.

Initial analysis using the IRIS-RSS data (Conrath et al., 1984) yielded a greatly sub-solar He/H=0.017±0.012 (He/$H_2$=2×He/H). Subsequent reanalysis of the data employing IRIS alone gave He/H between 0.055 and 0.08 (Conrath and Gautier, 2000). The authors emphasize, however, the retrieval of He/H is non-unique, but strongly suggest a value significantly greater than the earlier result that was based on the combined IRIS-RSS approach. For the purpose of this chapter, we take an average of the range of Saturn's He/H of 0.055-0.08, and express it as 0.0675±0.0125 (Table 2.1), but with the caveat that the value could well change following detailed analysis of the Cassini CIRS data and, especially, future in situ measurements at Saturn, as did Jupiter's He/$H_2$ following in situ measurements by the Galileo probe compared to the value derived from Voyager remote sensing observations. Current estimate of He/H in Saturn's upper troposphere is about 0.7× solar compared to Jupiter's 0.8× solar. The sub-solar He/$H_2$ in the troposperes of Jupiter and Saturn presumably results from the removal of some fraction of helium vapor through condensation as liquid at 1-2 megabar pressure in the interiors of these planets, followed by separation of helium droplets from metallic hydrogen. The severe depletion of Ne observed by the Galileo probe (Table 2.1) in Jupiter is excellent evidence of the helium-hydrogen immiscibility layer, as helium droplets absorb neon vapor, separate from hydrogen, rain toward the core, thus resulting in the depletion of helium and neon in the upper troposphere (Roulston and Stevenson, 1995; Wilson and Militzer, 2010). Models predict that the cooler interior of Saturn is expected to result in a greater degree of helium condensation and therefore a tropospheric He/$H_2$ ratio lower for Saturn than for Jupiter. Although the central value for Saturn is smaller than Jupiter's, the large uncertainty of Saturn's result does not provide a definite answer. Helium differentiation in Saturn's interior is invoked also as a way to explain the planet's large energy balance (Conrath et al., 1989). Without such chemical differentiation, models predict the heat flux excess at Saturn about three times lower than observed (Grossman et al., 1980), but the equation of state for the high pressure, high temperature interior is uncertain so the modeled excess is not that well constrained (see chapter by Fortney et al. for additional details). Saturn and Jupiter both emit nearly twice the thermal radiation compared to the radiation absorbed from the sun. Whereas the release of heat of accretion from conversion of the gravitational potential energy as these planets cool and contract over time accounts for a good fraction of the energy balance of Jupiter, helium differentiation may play a significant role at Saturn. Since helium is denser than hydrogen, gravitational potential energy available for conversion to heat increases as helium raindrops begin to separate from hydrogen and precipitate upon reaching centimeter size. In summary, there are indications that helium is depleted relative to solar in Saturn's troposphere, but the extent of such depletion will continue to be a subject of debate until precise in situ measurements

---

*Footnotes continued from the previous page*

[g] Fletcher et al. (2009a) derived global $PH_3$ mole fractions of 1.86±0.1 ppm and 6.41±0.42 ppm, respectively, in the upper tropospheres of Jupiter and Saturn from an analysis of the mid-IR emission measured by the Cassini Composite Infrared Spectrometer (CIRS).

[h] Conrath and Gautier (2000) give a range of 0.11-0.16 for the He/$H_2$ mole fraction from re-analysis of the Voyager IRIS data at Saturn, but the result is tentative. We use an average He/H=0.0675 for the purpose of calculating the ratios of other elements relative to hydrogen in Saturn.

[i] Fletcher et al. (2009b) report mole fraction of $CH_4$=4.7±0.2×$10^{-3}$ from an analysis of the CIRS data.

[j] Fletcher et al. (2011), using VIMS data giving an ammonia mole fraction, $f_{NH3}$, in the 1-3 bar region that is 140±50 ppm (scattering), 200±80 ppm (non-scattering) and rises to 300-500 ppm at the equator. If the maximum in ammonia measured at the equator (300-500 ppm, or 400±100 ppm) represents deep atmospheric $NH_3$, the corresponding $NH_3$/H = 2.27±0.6×$10^{-4}$.

[k] Briggs and Sackett (1989), using the VLA and the Arecibo microwave and radio data. The authors reported 10× solar $H_2S$, using solar S/H = 1.88×$10^{-5}$ from then current listing (Cameron, 1982). The S/H result in questionable (see text).

[l] Protosolar values based on present-day solar photospheric values of Lodders et al. (2009, table 4). The proto-solar abundances are calculated from the present-day values using the following corrections: +0.061 dex for He and +0.053 dex for all other elements.

can be made. In this regard, the final proximal orbits of Cassini in September 2017 are promising for the measurement of helium by the Ion and Neutral Mass Spectrometer down to ~1700 km or ≤0.1 nanobar (S. Edgington, personal comm., 2015), which is above Saturn's homopause level (1000-1100 km, or ~10-100 nanobar; Atreya, 1986, Strobel et al., this book), and perhaps deeper in the final trajectory when the spacecraft plunges into Saturn. Extrapolation to well-mixed troposphere would be model dependent even if the homopause level could be derived independently from the Cassini occultation data in the proximal orbits. Hence, precise helium abundance measurement directly in the well-mixed troposphere will still be essential, and that can only be done from an entry probe.

The nitrogen elemental abundance in Saturn is obtained from Saturn's principal nitrogen-bearing molecule, $NH_3$. From an analysis of the Cassini Visual and Infrared Mapping Spectrometer (VIMS) data, Fletcher et al. (2011) derive an ammonia mole fraction, $f_{NH3}$, in the 1-3 bar region that is 140±50 ppm (scattering), 200±80 ppm (non-scattering) and rising to 300-500 ppm at the equator. If we assume that maximum in ammonia measured at the equator (300-500 ppm, taken as $4\pm1\times10^{-4}$ here) represents also the $NH_3$ mole fraction in Saturn's deep well-mixed troposphere, then the corresponding $NH_3/H = 2.27\pm0.6\times10^{-4}$. That would imply an N/H enrichment of about 3× solar at Saturn, in contrast to Jupiter's roughly 5× solar. Previously, de Pater and Massie (1985) also found a 3× solar enhancement in Saturn's N/H in the 3-bar region, based on the VLA observations. The VLA and the Cassini RADAR 2.2 cm data (Laraia et al., 2013) also show that ammonia is subsaturated down to several bars, which most likely results from the loss of $NH_3$ in the lower clouds of $NH_4SH$ (or another form such as $(NH_4)_2S$) at ≥5 bars and the $NH_3-H_2O$ (aqueous-ammonia) solution cloud between approximately 10 and 20 bars depending on the enhancement of O/H ($H_2O$) above solar (Atreya et al., 1999; Atreya and Wong, 2005; see also Section 2.6 and Figure 2.9 therein). Whether the above 3× solar N/H in the 3-bar region is representative of the true nitrogen elemental ratio in Saturn's deep well-mixed troposphere is presently an open question, as the infrared or radio data can neither confirm nor rule it out. Unlike Saturn, there is no such ambiguity in the determination of Jupiter's N/H since direct in situ measurements of $NH_3$ could be made by the Galileo probe mass spectrometer (GPMS; Niemann et al., 1998) down to 21 bars, which is well below the $NH_3$ condensation level of 0.5-1 bar. Independently, $NH_3$ was derived also by analyzing the attenuation of the Galileo probe-to-orbiter radio communication signal (L-band at 1387 MHz or 21.6 cm) by ammonia in Jupiter's atmosphere (Folkner et al., 1998). $NH_3$ from the two sets of data agree to within 20%, with tighter constraints coming from the radio attenuation data, which yields N/H = 5.40±0.68× solar (Table 2.1). The Galileo probe value is likely representative of the global N/H in Jupiter, as the measurements were done well below any possible traps of ammonia, including condensation clouds of $NH_3$, $NH_4SH$ and $NH_3-H_2O$, with a caveat, however, that the probe region was dry so the final verdict will come from the Juno microwave radiometer measurements. The N/H value of Jupiter from Galileo is about twice the N/H in Saturn at 3 bars.

Sulfur is sequestered largely in the $H_2S$ gas in the atmospheres of Jupiter and Saturn. Whereas Jupiter's $H_2S$ could be measured directly and precisely in situ by the Galileo probe (Table 2.1), it was derived indirectly at Saturn by fitting the VLA and Arecibo microwave and radio data to assumed $NH_3$ abundances (Briggs and Sackett, 1989). Although direct microwave absorption by $H_2S$ could not be measured in these observations, they deduced $H_2S$ by analyzing $NH_3$ whose abundance is controlled to some extent by $H_2S$ since models predict it would remove a portion of the $NH_3$ vapor via the formation of an $NH_4SH$ cloud below. Using the then available solar $S/H=1.88\times10^{-5}$ (Cameron, 1982), they derived a ten times solar enrichment of S/H in Saturn's atmosphere, which translates into 12-13 times solar S/H using current solar S/H values, or about four times the value determined by the Galileo probe in Jupiter (Table 2.1). It is important to add a caveat, however. Whereas the Jupiter result comes from direct, in situ measurement of $H_2S$, the above result for Saturn is highly model-dependent, as it depends on the assumption of the formation of purported $NH_4SH$ cloud whose thermochemical properties are poorly constrained. Since sulfur is a key heavy element in the models of Saturn's formation, a fresh set of data on Saturn's $H_2S$ are warranted.

We list P/H in Table 2.1, but add a caveat that it may not represent the true P/H value in the deep well-mixed atmospheres of Jupiter or Saturn. This is because $PH_3$, the principal reservoir of phosphorus in the atmospheres of Jupiter and Saturn, is a disequilibrium species that is thermochemically stable in the deep atmosphere at pressures of about one thousand bars where the temperature is ~1000 K or greater (Fegley and Prinn, 1985; Visscher and Fegley, 2005), but it could only be measured in the upper troposphere/lower stratosphere. As $PH_3$ is dredged up from deep in the atmosphere to the upper atmosphere, it may potentially undergo loss due to oxidation to $P_4O_6$ by water vapor and solution in any water clouds along the way, or by other chemical reactions. Thus, the P/H ratio deduced from observations of $PH_3$ of Saturn and Jupiter in the upper atmosphere may represent a lower limit to the P/H ratio in their deep well-mixed atmosphere. Hence, the P/H values listed in Table 2.1 should not automatically be taken as a good proxy for the enrichment of other heavy elements not yet measured in Jupiter or Saturn. On the other hand, disequilibrium species such as $PH_3$, $GeH_4$, $AsH_3$ and CO are excellent tracers of the strength of convective mixing in the deep atmospheres of Saturn and Jupiter, and some could potentially be exploited to yield also a rough estimate of the deep water abundance.

Oxygen is arguably the most crucial of all heavy elements for constraining the formation models of Jupiter and Saturn. This is because in the reducing environments of the giant planets, oxygen is predominantly sequestered in water, which was presumably the original carrier of the heavy elements that formed the core and made it possible to accrete gas and complete the planet formation. [CO is another oxygen bearing species, but is million times less abundant than water.] Yet, the deep well-mixed abundance of water, hence O/H, remains a mystery. In the case of Jupiter, the Galileo probe entered an anomalously dry region known as a five-micron hot spot. In this "Sahara Desert of Jupiter", water was found to be severely depleted (Niemann et al., 1998; Atreya et al., 1999, 2003). Although the probe mass spectrometer measured water vapor down to 21 bars, i.e. well below the expected condensation level of $H_2O$ between 5-10 bars, it was still sub-solar at that level (Table 2.1), but rising. The determination of Jupiter's water abundance must await the Juno microwave radiometer observations in 2016-2017. No measurements of water vapor are available in Saturn's troposphere, however. Presence of water in Saturn's atmosphere is inferred indirectly from observations of visible lightning by Cassini's imaging spectrometer where lightning storm was predicted by Cassini's radio observations (Dyudina et al., 2010). Broadband clear filter observations showed visible lightning at ~35°S on the nightside in 2009 (Dyudina et al., 2010) and in blue wavelengths only on the dayside in the 2010-2011 giant lightning storm at ~35°N (Dyudina et al., 2013). These authors conjecture that a 5-10 times enhancement of water over solar can explain Saturn's lower occurrence rate of moist convection, an indicator of lightning, compared to Jupiter (Dyudina et al., 2010). Similarly, using thermodynamic arguments Li and Ingersoll (2015) conclude that Saturn's quasi-periodic giant storms recurring every few decades result from interaction between moist convection and radiative cooling above the water cloud base, provided that the tropospheric water vapor abundance is 1% or greater, i.e. O/H ≥10× solar. Such an enrichment in O/H would result in a droplet cloud of $NH_3$-$H_2O$ at ~20-bar level at Saturn (Atreya and Wong, 2005; see also Section 2.6 and Figure 2.9 therein). Although direct measurements of Saturn's well-mixed water may have to wait for future missions, as discussed in section 2.5, the recent discoveries of hot giant exoplanets and a Saturn-analog exoplanet are making it possible to measure $H_2O$ abundances in their atmospheres and in turn informing possible $H_2O$ abundances in solar system giant planets.

Highly precise measurements of methane in the atmosphere of Saturn have been carried out with Cassini's composite infrared spectrometer (CIRS) instrument (Flasar et al., 2005), which yield a mole fraction of $CH_4 = 4.7\pm0.2\times10^{-3}$ (Fletcher et al., 2009b). This results in a robust determination of the C/H ratio in Saturn (about twice the Jupiter value) that can be compared with rather imprecise but definitely higher estimates of C/H in Uranus and Neptune, as a way of constraining the giant planet formation scenarios.

Heavy noble gases, Ne, Ar, Kr and Xe, have been measured only in Jupiter's atmosphere (Table 2.1), since they can only be detected in situ by an entry probe, not by remote sensing. As noble gases are chemically inert, their abundances are unaffected by chemistry and condensation processes that control $NH_3$, $H_2S$, $H_2O$ and $PH_3$. Thus, the heavy noble gas enrichments are expected to be the same everywhere in the atmosphere. At Jupiter, with the exception of neon, they range from a factor of 2-3 times solar within the range of uncertainty of their planetary measurements and the solar values (Table 2.1). As neon dissolves in liquid helium, it is removed along with helium, which condenses in the 3 megabar region in Jupiter's interior, and is thus found depleted at observable shallow tropospheric levels (Wilson and Militzer, 2010). At Saturn, neon is expected to meet the same fate.

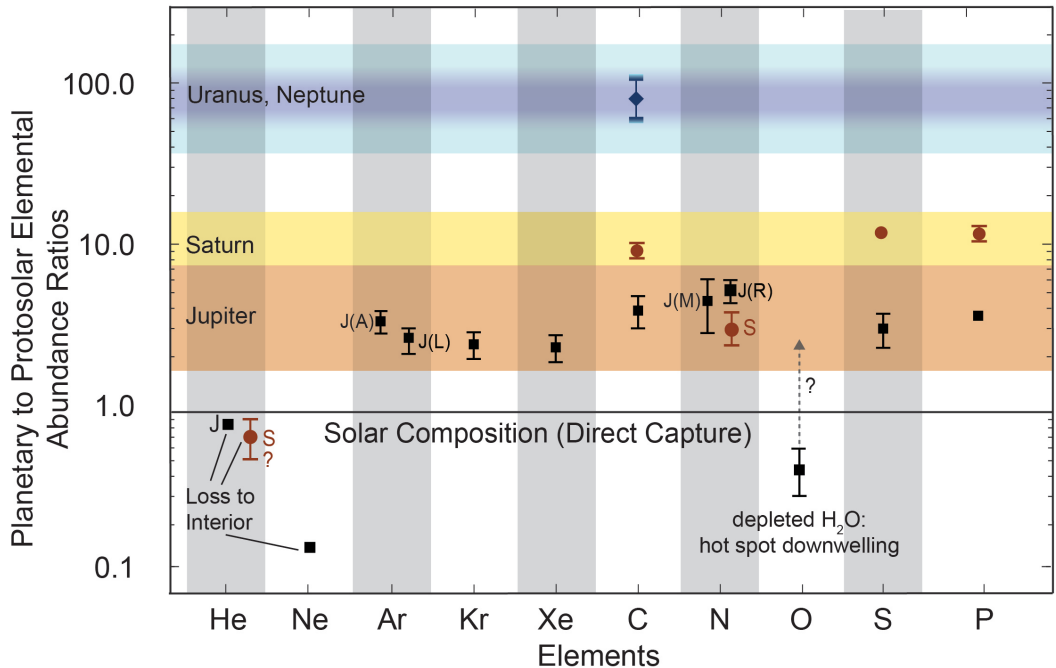

**Figure 2.1.** Abundances of key elements in the atmospheres of Saturn (brown dots, and label S) and Jupiter (black squares) relative to *protosolar* values derived from the present-day photospheric values of Asplund et al. (2009). Only C/H is presently determined for Uranus and Neptune, though poorly; its best estimate from earth-based observations is shown. The values are listed in Table 2.1. All values are ratioed to H (multiply by 2 for ratio to $H_2$). Direct gravitational capture would result in solar composition, i.e. no volatile enrichment, hence they would all fall on the horizontal line (normalized to solar) in the middle of the figure. Only He, C, N, S and P have been determined for Saturn, but only C/H is robust for the well-mixed atmosphere (see text). The Jupiter values are from the Galileo probe mass spectrometer (GPMS), except for N/H that was measured by both the GPMS [J(M)] and from attenuation of the probe radio signal through the atmosphere [J(R)]. For Ar, enrichments using both Asplund et al. [J(A)] and Lodders et al. [J(L)] solar values are shown. O/H is sub-solar in the very dry entry site of the Galileo Probe at Jupiter, but was still on the rise at the deepest level probed. Helium is depleted in the shallow troposphere due to condensation and differentiation in the planetary interior. Ne was also depleted in Jupiter as neon vapor dissolves in helium droplets.

Figure 2.1 shows the enrichment factors of the heavy elements and He in the atmospheres of Saturn and Jupiter relative to their protosolar values (all ratioed to H). Here we use Asplund et al. (2009) compilation of photospheric elemental abundances (their table 1), as they represent an improvement over previous conventional standards (e.g. Anders and Grevesse, 1989; Grevesse et al., 2005, 2007) and result from the use of 3D hydrodynamic model of the solar atmosphere, nonlocal thermodynamic equilibrium effects, and improved atomic and molecular data. The photospheric values are then converted to protosolar elemental abundance (see table footnote). The latter account for the effects of diffusion at the bottom of the convective zone on the chemical composition of the photosphere, together with the effects of gravitational settling and radiative accelerations. According to Asplund et al. (2009), the protosolar metal abundances relative to hydrogen can be obtained from the present day values increased by +0.04 dex, i.e. ~11%, with an uncertainty of ±0.01 dex; the effect of diffusion on He is very slightly larger: +0.05 dex (±0.01). Lodders et al. (2009) suggest a slightly larger correction of +0.061 dex for He and +0.053 dex for all other elements. Previously, Grevesse et al. (2005, 2007) used the same protosolar correction of +0.05 dex for all elements (dex stands for "decimal exponent", so that 1 dex=10; it is a commonly used unit in astrophysics).

Figure 2.1 is based on protosolar correction to Asplund et al. (2009) photospheric abundances, while Table 2.1 lists planetary elemental enrichment factors also for Lodders et al. (2009) protosolar values. Whereas the difference between the enrichment factors based on Asplund et al. and Lodders et al. values is at most 10-15% for most elements, Asplund et al. estimate nearly 30% greater enrichment for Ar/H compared to Lodders et al. (Table 2.1).

The difference in Jupiter's Ar enrichment factors based on Asplund et al. (2009) and Lodders et al. (2009) can be traced back largely to the choice of O/H employed by the two sets of authors. Because of their high excitation potentials, noble gases do not have photospheric spectral features; hence their solar abundances are derived indirectly. Asplund et al. (2009) infer solar Ar/H following the same procedure as Lodders (2008), i.e. by using, amongst other things, the Ar/O data from the solar wind, solar flares and the solar energetic particles, but employing their own photospheric abundances of O/H that have a somewhat lower uncertainty than Lodders et al. (2009). This accounts for much of the abovementioned 30% difference in Jupiter's Ar/H enrichment factor. Nevertheless, within the range of uncertainty of Jupiter's argon abundance and the dispersion in the solar values, the Ar/H enrichment in Jupiter relative to the solar Ar/H is nearly the same whether one uses Asplund et al. (2009) or Lodders et al. (2009) solar Ar/H. We show both results in Figure 2.1. A word of caution about oxygen, which is used by above authors as a proxy for deriving the solar Ar/H, is in order, however, as explained below.

Ever since concerted efforts were made to determine the solar elemental abundances, particular attention has been paid to oxygen, as oxygen is the most abundant element that was not created in the Big Bang, and third only to H and He that were created in the Big Bang. Furthermore, the principal reservoir of oxygen in Saturn and Jupiter, $H_2O$, was presumably the original carrier of the heavy elements to these planets. Thus, oxygen is centrally important to the question of origin of all things. Yet, its abundance in the sun has been revised constantly. As illustrated in Figure 2.2 the solar O/H values have gyrated up and down several times in the past four decades, starting with the classic work of Cameron (1973) to the present. The highest solar O/H value is the one recommended by Anders and Grevesse (1989), which remained the standard for a good fifteen years, only to be revised downward by nearly a factor of two in 2005 (Grevesse et al., 2005), and creeping up a bit since then. Not surprisingly, the solar Ar/H, also plotted in Figure 2.2, shows the same trend as O/H over time, though they are not completely proportional to each other nor are they expected to be. Thus, one needs to be vigilant about changes in the photospheric abundance of oxygen and other elements such as argon that use oxygen as a reference.

In summary, the most robust elemental abundance determined to date in Saturn is that of carbon. At 9× solar, Saturn's C/H is a little over twice the C/H ratio in Jupiter. This is consistent with the core accretion model of giant planet formation, according to which progressively increasing elemental abundance ratios are expected from Jupiter to Neptune. Carbon is the only heavy element ever determined for all four giant planets (Figure 2.1), and indeed it is found to increase from 4× solar in Jupiter to 9× solar in Saturn, rising to 80(±20)× solar or greater in both Uranus (Sromovsky et al., 2011; E. Karkoschka and K. Baines personal communication, 2015) and Neptune (Karkoschka and Tomasko, 2011), using the current solar C/H from Table 1. The same trend is also seen in the S/H ratio of Saturn compared to Jupiter, except for a four-fold increase from Jupiter to Saturn, but Saturn's S/H is less secure as discussed above. The difference in the relative changes of C/H and S/H is worth noting, but caution should be exercised to not over interpret it. This is because $H_2S$ is a thermochemically condensible volatile in the gas giants unlike $CH_4$. Saturn's S/H would benefit greatly from fresh set of modern data. Similar four-fold increase is seen also in the P/H ratio in Saturn compared to Jupiter, and the relative change may be valid if the disequilibrium species $PH_3$ meets a similar fate in the tropospheres of Saturn and Jupiter. On the other hand, the observed 3× solar N/H ratio in Saturn seems puzzling, as it is about a factor of two less, not more, than Jupiter's N/H, contrary to the predictions of conventional formation models. However, the present data on Saturn's $NH_3$ in the 3 bar region do not rule out much greater ammonia abundance in the deep well-mixed atmosphere of Saturn as discussed above. Presence of water is inferred in Saturn's troposphere indirectly from localized lightning observations, but no firm conclusions can be drawn from it on the global O/H ratio in Saturn. The Juno spacecraft is designed to measure and map water to several hundred bars in Jupiter's troposphere, which will provide a definitive answer on Jupiter's O/H ratio. In Jupiter at least, for which data are available for most of the heavy elements, except for O/H, it is striking that the heavy noble gases, Ar, Kr and Xe all display similar enrichment over solar by a factor of 2-3 (or, 2-2.5 with Lodders solar values, Table 2.1), whereas enrichment of non noble gas elements, carbon, nitrogen and sulfur, is greater, ranging from 4-6. [Regarding S/H, from their clathrate hydrate model Gautier et al. (2001) calculate an S/H enrichment in Jupiter, which is twice the value measured by the Galileo probe (Table 2.1), or ~6× solar, and attribute the lower measured value to the loss of $H_2S$ in troilite (FeS) in the inner solar nebula.] Though it may seem tempting and convenient to lump them all together and suggest that the heavy elements in Jupiter are enriched uniformly by a factor of 4±2 relative to their solar abundances, we advise caution.

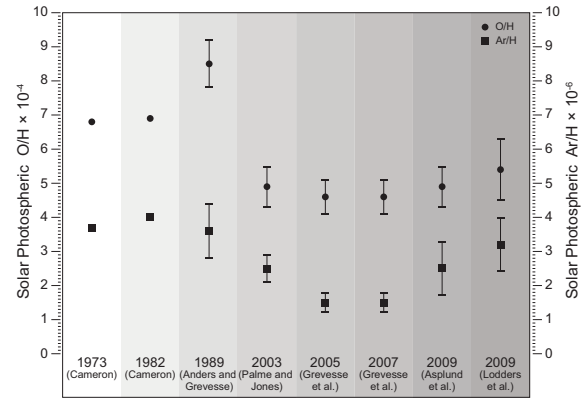

**Figure 2.2.** Time history of the solar photospheric O/H and Ar/H, showing only the major milestones. Although Ar/H shows the same trend as O/H, they do not track each other exactly. The solar photospheric values for O/H ($\times 10^{-4}$) and Ar/H ($\times 10^{-6}$) plotted here are, respectively, 6.8 and 3.7 (Cameron, 1973), 6.9 and 4 (Cameron, 1982), 8.5±0.7 and 3.6±0.8 (Anders and Grevesse, 1989), 4.9±0.6 and 2.5±0.4 (Palme and Jones, 2003), 4.6±0.5 and 1.5±0.3 (Grevesse et al., 2005, 2007), 4.9±0.6 and 2.5±0.8 (Asplund et al., 2009), and 5.4±0.9 and 3.2±0.8 (Lodders et al., 2009).

The differences between the enrichments of the heavy noble gases and those of the non noble gas heavy elements are apparently real, and may indicate two distinct populations arising from differences in the way noble gases were delivered (see also Section 2.4.3). Robust measurements of the same set of heavy elements at Saturn as Jupiter are crucial to determine whether they have solar composition, as proposed by Owen and Encrenaz (2006), which will in turn have a bearing on the models of the origin, nature and delivery of the Saturn forming planetesimals. Similar efforts are now underway to measure key elemental abundances, particularly of O and C, in the atmospheres of giant exoplanets and in using them to constrain formation conditions of exoplanetary systems (see section 2.5).

### 2.2.2 Isotopic composition of Saturn's atmosphere and comparison to Jupiter

Isotope ratios provide an insight into the conditions prevailing at the time of formation of the solar system and even early in the beginning of the universe. The giant planets and the terrestrial planets formed from much of the same initial inventory of material in the primordial solar nebula. Thus, the stable gas isotope ratios originally were the same in all planets. Abiotic fractionation of isotopes can occur due to escape of gases to space, loss to surface, phase change, or photochemistry. Indeed, fractionation of various stable gas isotopes has been found in the atmospheres of comparatively small solar system objects including Venus, Earth, Mars and Titan (e.g. von Zahn et al., 1983; Niemann et al., 2010; Atreya et al., 2013; Webster et al., 2013; Mahaffy et al., 2014), and attributed mainly to the loss of their volatiles to space over geologic time. On the other hand, the sheer mass of the giant planets, in particular Jupiter and Saturn, does not permit loss of volatiles either by thermal, charged particle or other processes, hence their original isotopic ratios of elements are expected to be preserved for all practical purposes. Thus, their present atmospheric isotope ratios should, in principle, also represent protosolar values.

Only a handful of the isotopes have been measured in Saturn's atmosphere: $^{13}C/^{12}C$, D/H, and an upper limit on $^{15}N/^{14}N$. In the atmosphere of Jupiter, $^{3}He/^{4}He$, $^{36}Ar/^{38}Ar$, all isotopes of Xe except for $^{124}Xe$ and $^{126}Xe$ that together comprise 0.2% of total xenon in the sun, have been measured in addition to $^{13}C/^{12}C$, D/H, $^{15}N/^{14}N$. The measurement of noble gas isotopes in Jupiter was facilitated by in situ measurements with a mass spectrometer on the Galileo probe (GPMS). The isotope ratios for the atmosphere of Saturn and Jupiter are listed in Table 2.2. The helium, carbon and xenon isotope ratios of Jupiter are nearly identical to the solar values, as expected.

The hydrogen isotope ratio, D/H, in Jupiter and Saturn is important for understanding the very beginnings of the universe and galactic evolution. Deuterium was formed following the Big Bang, but has been declining ever since because of its destruction in the stars far outweighs any creation. Thus, the D/H ratio in Jupiter and Saturn represents the protosolar value of D/H in the sun, in which it cannot be measured directly today. The value derived by the GPMS in Jupiter's atmosphere was thus the first measurement of the protosolar D/H ratio (Mahaffy et al., 1998). The result is in agreement with the D/H measurements done later with the short wavelength spectrometer on the Infrared Space Observatory (ISO, Lellouch et al., 2001) and theoretical estimates (Table 2.2). This gives confidence in the D/H ratio measured by ISO in Saturn's atmosphere. Within the range of uncertainty, Saturn's D/H ratio is similar to that in Jupiter.

The nitrogen isotope ratio was measured in Jupiter's atmosphere by the Galileo probe mass spectrometer (Owen et al., 2001), and represented the first measurement of the protosolar $^{15}N/^{14}N$ ratio. The value in the sun is now available from the Genesis measurements (Marty et al., 2011) and is identical to the GPMS result for Jupiter. The ISO data give a slightly lower $^{15}N/^{14}N$, probably resulting from isotope fractionation below the ammonia clouds to which the ISO data apply. Note, however, $^{15}N/^{14}N$ from ISO has large uncertainties that can easily envelope the GPMS result. Unlike Jupiter, only an upper limit on the $^{15}N/^{14}N$ ratio in Saturn's atmosphere is available. Using the Texas Echelon cross Echelle Spectrograph (TEXES) on NASA's Infrared Telescope Facility (IRTF), Fletcher et al. (2014) observed spectral features of $^{14}$NH$_3$ and $^{15}$NH$_3$ in 900 cm$^{-1}$ and 960 cm$^{-1}$, and derived an upper limit on the $^{15}$N/$^{14}$N ratio of 2×10$^{-3}$ for the 900 cm$^{-1}$ channel and 2.8×10$^{-3}$ for the 960 cm$^{-1}$ channel. Though these values fall in the range of Jupiter's $^{15}$N/$^{14}$N ratio, in the absence of actual measurement they represent only upper limits of $^{15}$N/$^{14}$N in Saturn's atmosphere. In Figure 2.3, we show the best available data on this important ratio in the sun, interstellar medium, Jupiter, Saturn, and comets (from CN, HCN and NH$_2$), which represent the original reservoirs of nitrogen (left panel, labeled "Primordial"), and in N$_2$ of the terrestrial planets and Titan, where nitrogen is secondary (right panel, labeled "Secondary"). The corresponding nitrogen isotope ratios are listed in Table 2.3. Nitrogen isotope fractionation is clearly evident in the terrestrial bodies. The lighter isotope floats up to the top of the atmosphere and escapes preferentially, leading to the build-up of the heavier isotope.

**Table 2.2.** Elemental isotopic ratios in the sun, Jupiter and Saturn

| Elements | Sun | Jupiter | Saturn |
| --- | --- | --- | --- |
| $^{13}C/^{12}C$ | $0.0112^{(a)}$ | $0.0108\pm0.0005^{(i)}$ | $0.0109\pm0.001^{(o)}$ |
| $^{15}N/^{14}N$ | $2.27\pm0.08\,10^{-3(b)}$ | $(2.3\pm0.03)\times10^{-3}$ (0.8–2.8 bar)$^{(j)}$ $1.9(+0.9,-1.0)\times10^{-3}$ (0.2–1.0 bar)$^{(k)}$ | $<2.0\times10^{-3(p)}$ (900 cm$^{-1}$ channel) $<2.8\times10^{-3(p)}$ (960 cm$^{-1}$ channel) |
| $^{36}Ar/^{38}Ar$ | $5.5\pm0.0^{(c)}$ | $5.6\pm0.25^{(l)}$ | |
| $^{136}Xe/Xe$ | $0.0795^{(a)}$ | $0.076\pm0.009^{(l)}$ | |
| $^{134}Xe/Xe$ | $0.0979^{(a)}$ | $0.091\pm0.007^{(l)}$ | |
| $^{132}Xe/Xe$ | $0.2651^{(a)}$ | $0.290\pm0.020^{(l)}$ | |
| $^{131}Xe/Xe$ | $0.2169^{(a)}$ | $0.203\pm0.018^{(l)}$ | |
| $^{130}Xe/Xe$ | $0.0438^{(a)}$ | $0.038\pm0.005^{(l)}$ | |
| $^{129}Xe/Xe$ | $0.2725^{(a)}$ | $0.285\pm0.021^{(l)}$ | |
| $^{128}Xe/Xe$ | $0.0220^{(a)}$ | $0.018\pm0.002^{(l)}$ | |
| $^{20}Ne/^{22}Ne$ | $13.6^{(a)}$ | $13\pm2^{(l)}$ | |
| $^{3}He/^{4}He$ | $1.66\times10^{-4\,(a)}$ $(1.5\pm0.3)\times10^{-4}$ (meteoritic)$^{(d,e,f,g)}$ | $(1.66\pm0.05)\times10^{-4(m)}$ | |
| D/H | $(2.0\pm0.5)\times10^{-5(a)}$ $(2.1\pm0.5)\times10^{-5(h)}$ *protosolar* values | $(2.6\pm0.7)\times10^{-5(m)}$ $(2.25\pm0.35)\times10^{-5(n)}$ | $1.7(+0.75,-0.45)\times10^{-5(n)}$ |

$^{(a)}$Asplund et al. (2009), updated from Rosman and Taylor (1998); $^{(b)}$Marty et al. (2011) from Genesis; $^{(c)}$Vogel et al. (2011); $^{(d)}$Black (1972); $^{(e)}$Eberhardt (1974); $^{(f)}$Geiss and Reeves (1972); $^{(g)}$Geiss (1993); $^{(h)}$Geiss and Gloeckler (1998); $^{(i)}$Niemann et al. (1998); $^{(j)}$Owen *et al.* (2001), from Galileo probe mass spectrometer (GPMS) in situ measurements, largely below the NH$_3$ condensation level; $^{(k)}$Fouchet *et al.,* (2000), from ISO infrared remote sensing measurements, largely above the NH$_3$ condensation level; $^{(l)}$Mahaffy *et al.* (2000), normalized to 1.0 for xenon isotopes measured, only $^{126}$Xe and $^{124}$Xe, which together make up 0.2% of the total xenon in the sun, could not be measured by the GPMS in Jupiter, and the xenon error bars are with respect to the ratio of each isotope to its non-radiogenic terrestrial value; $^{(m)}$Mahaffy et al. (1998), from GPMS; $^{(n)}$Lellouch et al., (2001) from ISO; $^{(o)}$Fletcher et al. (2009b); $^{(p)}$Fletcher et al. (2014).

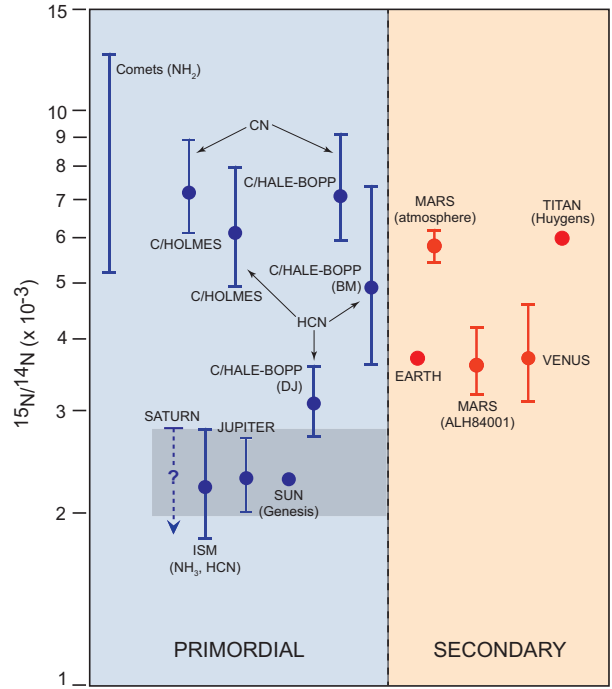

**Figure 2.3.** A comparison of the current nitrogen isotope ratios in primordial (sun, ISM, Jupiter, Saturn and comets; left panel) and secondary (Venus, Earth, Mars and Titan; right panel) reservoirs. The values for the secondary reservoirs illustrate loss of nitrogen from these objects over geologic time. A subset of the $^{15}N/^{14}N$ ratios based on then available data can be found in Owen et al. (2001) and Atreya et al. (2009). References for current ratios are listed in Table 2.3. For Mars, the value in ALH84001 is also shown, and represents $^{15}N/^{14}N$ for solid Mars, which is nearly the same as for the atmosphere of Earth, as expected, considering that these planets presumably acquired their original building material from the same source. The substantially higher $^{15}N/^{14}N$ in Mars atmosphere today compared to Earth's atmosphere is the consequence of thermal and solar wind induced escape of nitrogen from (lighter and non-magnetic) Mars over geologic time. Higher $^{15}N/^{14}N$ in Titan's present atmosphere compared to Earth's atmosphere may reflect the value in its building blocks, coupled with atmospheric loss over time. In comets, similar value of $^{15}N/^{14}N$ has been measured in CN of a dozen comets, and in CN and HCN of comets Hale Bopp and Holmes. However, the $^{15}N/^{14}N$ in HCN in the the latter two comets is substantially different from that in CN, and different observers (BM: Bockelée-Morvan et al., 2008, DJ: Jewitt et al., 1997) report different values though all have large error bars. $^{15}N/^{14}N$ from $NH_2$ in a number of comets (Rousselot et al., 2014) is shown as a range, which is more appropriate than an average value because of the difficulty of accounting for solar continuum for each emission feature of $NH_2$ according to the authors. The reader is referred to Atreya et al. (2009) for additional discussion on the cometary $^{15}N/^{14}N$ and implications for Titan. For Saturn only upper limits from 900 cm$^{-1}$ and 960 cm$^{-1}$ spectral channels are available.

### 2.2.3 Saturn's interior

Saturn's interior may be probed indirectly through models and the measurement of the planet's mean density and gravitational moments (Fortney et al., this book) and measurement of the planet's dissipation factor (e.g. Remus et al., 2012). It has long been known that it is mostly made of hydrogen and helium, except for the presence of a central dense core. Detailed models show that in spite of its low global density of 0.688 g/cm$^3$, Saturn must contain a significant fraction of its mass as heavy elements: between about 12 and 28 $M_E$ (Nettelmann et al., 2013; Helled and Guillot, 2013), corresponding to a mass fraction $Z=0.13$ to $0.29$ or a global enrichment in heavy elements of 8.9 to 20 times the solar value.

In classical 3-layer models, most of the heavy elements are embedded in a central core. The solutions of Helled & Guillot (2013), assuming a well-defined central core and a homogeneous abundance of heavy elements in the envelope, indicate a core with a mass between 10 and 20 $M_E$ and an envelope with 4 to 8 $M_E$ of heavy elements, corresponding to an enrichment of 4 to 8 times the solar value. The abundances of C, N and S bearing species in the atmosphere account for the lower limit of this range, meaning that the elusive O can be enriched only as much as C. As the "total" enrichment in heavy elements is constrained by the interior models, the addition of other species (e.g. silicates) into the envelope would mean less enrichment for others, which would make O even less enriched, implying a C/O ratio that is very likely to be supersolar in Saturn's atmosphere. Solutions by Nettelmann et al. (2013) add one degree of freedom, the possibility for the abundance of heavy elements to vary in the envelope. That leads to the possibility of even smaller core masses, but with a deep envelope that is enriched in heavy elements. Thus the picture that emerges is one in which a core is either well-defined or only partially mixed with the envelope, and an envelope that is significantly enriched in heavy elements, but not in the same way for all species. Accounting for material that may be partially mixed in the deep envelope, Saturn's core appears to have a mass that is consistent with that required by core-accretion models (e.g. Ikoma et al., 2001). The enrichment of the envelope is to be explained either by planetesimal impacts, or by upward mixing and/or core erosion. The former is traditionally difficult because the cross-section of a mature giant planet (i.e., when the planet has accreted its full mass and does not possess a circumplanetary disk anymore) is small. For example, simulations of impacts during the great heavy bombardment indicate that of an initial disk mass of 35 $M_E$, only between 0.05 and 0.1 $M_E$ hit Saturn (the values are about double for Jupiter, due to a larger focusing factor; Matter et al., 2009).

**Table 2.3.** Nitrogen isotope ratios in the solar system

| Objects | $^{14}N/^{15}N$ | $^{15}N/^{14}N$ (×10$^{-3}$) |
|---|---|---|
| Sun (protosolar)[a] | 441±5 | 2.27±0.03 |
| Jupiter[b] | 442±58 | 2.30±0.3 |
| Saturn[c] | < 357 | < 2.8 |
| Interstellar Medium (ISM)[d] | 450±98 | 2.2±0.5 |
| Comet Hale-Bopp (CN)[e] | 140±30 | 7.1(+2.0, -1.3) |
| Comet Hale-Bopp (CN)[f] | 140±35 | 7.1(+2.4, -1.4) |
| Comet Hale-Bopp (HCN)[g] | 323±46 | 3.1(+0.5, -0.4) |
| Comet Hale-Bopp (HCN)[f] | 205±70 | 4.9(+2.5, -1.3) |
| Comet Holmes (CN)[f] | 139±26 | 7.2(+1.7, -1.1) |
| Comet Holmes (HCN)[f] | 165±40 | 6.1(+1.9, -1.2) |
| Comets (NH$_2$)[h] | 80-190 | 5.26-12.5 |
| Earth | 272 | 3.68 |
| Venus[i] | 272±54 | 3.7(+0.9, -0.6) |
| Mars (atmosphere)[j] | 173±11 | 5.8±0.4 |
| Mars (solid body)[k] | 276.5±0.25 | 3.62 |
| Titan[l] | 167.7±0.6 | 6.0±0.02 |

[a]Marty et al. (2011), from Genesis sample analysis; [b]Owen et al. (2001), Galileo probe mass spectrometer; [c]Fletcher et al. (2014), IRTF; [d]Dahmen et al. (1995); [e]Arpigny et al. (2003); [f]Bockelée-Morvan et al. (2008); [g]Jewitt et al. (1997); [h]Rousselot et al. (2014), derived from emission lines of NH$_2$ in twelve comets between 2002 and 2013, and the authors state that the range in $^{14}N/^{15}N$ is probably more appropriate to use than the average value of 127, which does not account for uncertainties because of the difficulty in accurately subtracting the solar continuum for each region of interest; [i]Hoffman et al. (1979), Pioneer Venus; [j]Wong et al. (2013), from MSL; [k]Matthew and Marti (2001), from the oldest known martian meteorite, ALH84001 (4.1 Gyr old); [l]Niemann et al. (2010), Huygens-GCMS.

Core erosion is made possible from a physical point of view because of the miscibility of species in metallic hydrogen (Wilson and Militzer, 2010, 2012). However, while it is effective at Jupiter, Saturn's smaller envelope implies that only about 2 $M_E$ may be mixed upward from a massive central core, assuming a 10% efficiency of the process (Guillot et al., 2004). A higher efficiency, or more likely the upward mixing of an initially heavy-element rich primordial envelope could explain the planet's heavy element rich atmosphere. Variations in the elemental composition (such as those leading to a supersolar C/O ratio) could be explained by a selective retention of species (e.g. silicates, water) in the deeper regions.

## 2.3 Saturn's formation: hydrodynamical point of view

Standard models of Saturn's interior with a core surrounded by a hydrogen-helium envelope that is enriched in heavy elements fit well with picture of its formation by core accretion followed by the capture of the gas envelope from the protoplanetary disk. However, considerable uncertainties remain, both on the internal structure itself, and on formation models. To understand the end-to-end origin and evolution of Saturn, it is important to consider then the starting protosolar disk, the manner of formation and growth of the core, Saturn's circumplanetary disk, and any insight from the moons and rings. This section discusses each of these aspects from a hydrodynamical point of view.

### 2.3.1 Birth and evolution of the protosolar disk

Any model of Saturn formation must begin with the protoplanetary disk, or solar nebula, from which the gas and dust of Saturn were derived. Constraints on giant planet formation include the disk lifetime, elemental composition (specifically, C/H, O/H, etc.,) and the overall mass of the disk. A low opacity massive disk may fragment very early by disk instability, but we argued in Section 2.1 that the overall architecture of our solar system is not matched by such an event. Core accretion, then, is constrained to build Saturn within a plausible disk lifetime. The model of Dodson-Robinson et al. (2008) provides a particular example of the detailed specification of a solar nebula model needed to build Saturn in an acceptably short length of time.

### 2.3.2 Formation and growth of giant planet cores

*Core formation*
In the framework of the core accretion model, the first step is obviously to form a 10 Earth masses core. In the classical view, gravity is the dominant process, and kilometer sized planetesimals merge when they collide. In the end, a population of so-called oligarchs is produced, which accrete all the planetesimals within reach of their orbit (Kokubo and Ida, 1998). Their mass is then typically $0.05(r/1AU)^{0.75} M_E$.

Another model suggests that centimeter-size dust aggregates are concentrated by vortices in the gas up to the point where the concentration of solids becomes gravitationally unstable, leading possibly directly to the formation of solid bodies of hundreds of kilometers (Johansen et al., 2007; see Turner et al., 2014a for a review of turbulent processes).

It has been shown recently that such embryos are very efficient at accreting pebbles, i.e. cm sized aggregates moderately coupled to the gas (Lambrechts and Johansen, 2012; Morbidelli and Nesvorny, 2012). As such pebbles drift radially, nothing stops this growth, whose rate is exponential. Pebble accretion is to date the most promising mechanism to form a few Earth masses core within the lifetime of a protoplanetary disk. Furthermore, Lambrechts et al. (2014) show that pebble accretion naturally stops when the core becomes massive enough to carve a dip in the gas that stops the radial drift of pebbles (~20 $M_E$). The end of the accretion of solids by the core then triggers the onset of the runaway accretion of gas.

*Planet migration*
Cores and planets interact gravitationally with the gas disk. This leads to exchanges of energy and angular momentum, hence to a variation of the orbit of the planet. This is called planetary migration. Bodies of less than roughly 50 $M_E$ do not perturb the gas profile much and are in the type I migration regime (Ward, 1997). It has been shown in the last decade that type I migration can be directed inwards or outwards, depending on the thermodynamics of the gas disk (Paardekooper and Mellama, 2006; Paardekooper et al., 2010, 2011). Typically, migration would be directed inwards in the outer, optically thin regions of the disk, while it can be directed outwards in the inner, optically thick regions. This opens the possibility of convergent migration towards a zero-torque migration radius where bodies of few Earth masses should gather, and hopefully merge (Lyra et al. 2010, Cossou et al., 2013). In general, there are two such radii, whose locations depend on the disk structure (Bitsch et al., 2013). One is inside the snowline and vanishes when the accretion rate in the disk decreases, and one is beyond the snowline, moving from roughly 10 AU to 4 AU as the disk ages (Bitsch et al., 2014). In any case, this new vision of planet migration (see Baruteau et al., 2014 for a recent and complete review) opens the possibility to keep the core of a giant planet safe at the zero torque migration radius, instead of losing it into the star. It can then grow by accreting slowly its gas envelope.

When it is massive enough, the planet will open a gap in the gas disk (Papaloizou and Lin, 1984; Crida et al., 2006), and therefore leave the type I migration regime. Planets opening gaps are in the type II, slower mode of migration, in which they follow roughly the viscous evolution of the disk (Lin and Papaloizou, 1986; Crida and Morbidelli, 2007; Dürmann and Kley 2015).

It should be noted that the migration of the Jupiter-Saturn pair is however more complex than that of a single giant planet. Jupiter and Saturn most likely enter in mean motion resonance, which can reverse their migration (Masset and Snellgrove, 2001). A fine tuning of the disk parameters allows the Jupiter and Saturn pair to avoid any significant migration in the protosolar nebula (Morbidelli and Crida, 2007). Another possibility is that Jupiter grew and migrated inwards first, then was caught up by Saturn, which made the pair migrate back outwards (Walsh et al., 2011). In this so-called "Grand Tack" scenario, the main asteroid belt is satisfactorily reproduced, and Jupiter's excursion sculpts the inner disk of embryos and planetesimals in a very favorable way for the formation of the terrestrial planets. It implies that Saturn came as close as about 2 AU from the Sun. Little room is left for gas accretion in this scenario as the final masses of Jupiter and Saturn are ideal for such a tack, but Saturn could have been half its present mass, gaining the rest on the way out.

In any case, an unavoidable consequence of migration is that Saturn most likely was in resonance with Jupiter, on a circular orbit ~8 AU from the Sun, when the protosolar nebula dissipated. It reached its final orbit ~650 millions years later, during a global dynamical instability among the giant planets, often referred to as the "Nice model" (Tsiganis et al., 2005).

### 2.3.3 Formation of Saturn and its circumplanetary disk

As discussed above and illustrated in Figure 2.4, giant planets open gaps in the protoplanetary disk. The neighborhood of their orbit is depleted, splitting the disk into an inner and an outer disk. While the width of the gap is set solely by the Hill radius of the planet, the depth of the gap increases with the planet mass and is also a smooth function of the viscosity and aspect ratio of the disk (Crida et al., 2006). Even for massive planets like Saturn or Jupiter, the opening of the gap does not terminate gas accretion. Indeed, as can be seen on

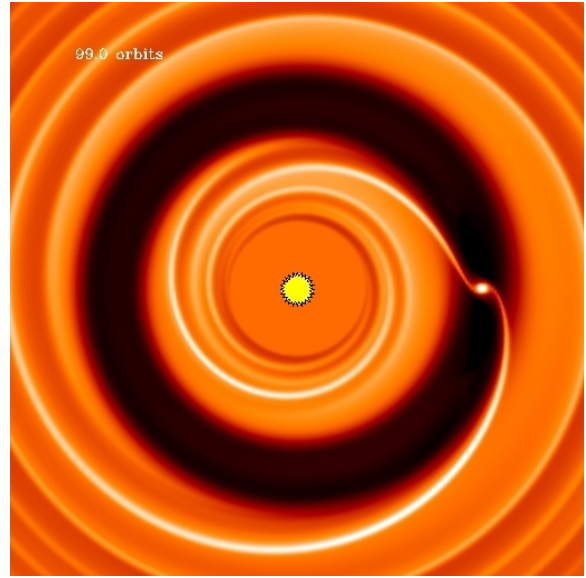

**Figure 2.4.** Gas density map from a 2D hydrodynamical simulation. Light color corresponds to high density, black to low density. The star is in the center of the image; the giant planet is on the right. The black annulus is the gap around the planetary orbit, and the white spot around the planet (not shown) is the CPD.

Figure 2.4, gas still flows towards the planet through the spiral wake. As a consequence, the final phase of runaway gas accretion corresponding to the collapse of the gas envelope (see Section 2.1) has no reason to end until a few Jupiter masses are reached. However, numerical simulations show that massive planets are capable of creating their own gas disk around them, inside the gap (Bate et al., 2003; Ayliffe and Bate, 2012). This circumplanetary disk (hereafter CPD) is poorly resolved in Figure 2.4, but has been studied in greater detail in other works.

The simulations reveal that the gas flow around a giant planet is 3D, and cannot be accurately modeled by a 2D simulation. Actually, most of the gas that reaches the CPD comes from a vertical direction, which is perpendicular to the orbital plane (Bate et al., 2003; Machida et al., 2008; Ayliffe and Bate, 2009a,b; Tanigawa et al., 2012; Szulagyi et al., 2014). An explanation for this unexpected flow pattern is given by Morbidelli et al. (2014): in the upper layers of the disk the gravitational force from the planet is weaker, and therefore, the gap tends to be narrower than in the midplane. Gas comes in, and being not supported by pressure, falls towards the midplane, where the planet ejects it back out of the gap, still in the midplane. Therefore, a meridional circulation pattern is created, as sketched by white arrows in Figure 2.5: gas ejected from the midplane by the planet expands vertically further away, and then slowly penetrates inside the gap from the upper layers, before falling back on the midplane. Note that such a full loop is much longer than an orbital period. Part of this vertical inflow falls on the CPD and the planet, contributing to the planet's growth.

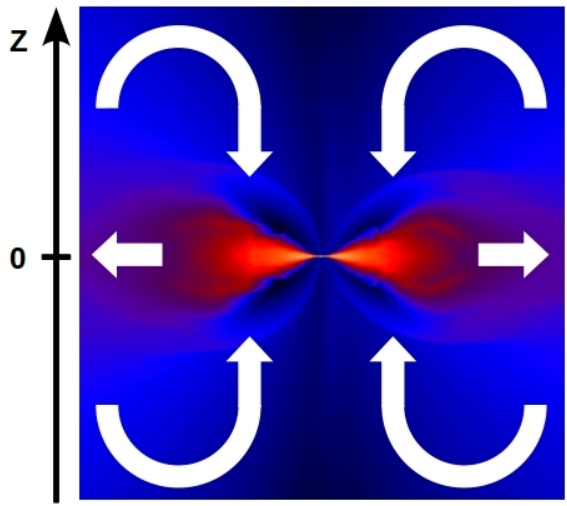

**Figure 2.5.** Vertical cut through the CPD of a Jupiter mass planet in a 3D simulation. The horizontal axis represents the distance to the star and the vertical axis is perpendicular to the orbital plane of the planet, with the planet in the center. The color corresponds to the gas density, and the arrows sketch the meridional circulation of the gas.

This may have strong implications on the nature and rate of the solids in this flow. By the time a giant planet forms, dust is supposed to have sedimented in the mid-plane of the disk, being only stirred by turbulence. Micrometric grains, well coupled to the gas, should follow the gas flow, but larger aggregates may well be unable to reach the planet and its CPD.

In isothermal simulations (the only ones available so far in the literature), gas falls at a supersonic speed onto the CPD and shocks at the surface of the latter. In the absence of viscosity in the CPD, gas should in principle reach its centrifugal radius and orbit around the planet forever. This opens the possibility that the CPD acts as a bottleneck for planet growth, and limits the final mass of the giant planets (Rivier et al., 2012). There are good reasons to think that the CPD is really inviscid (Turner et al., 2010, 2014b; Fujii et al., 2011, 2014). Hence, Szulagyi et al. (2014) have measured the different sources of angular momentum loss in the CPD (torque from the central star, contact with the infalling gas), using mesh refinement around the planet in a 3D global simulation, allowing for the gap to form accurately. They deduce a mass doubling time of the order of half a million years for a Jupiter mass planet. This is much slower than the standard, 1D model of Pollack et al. (1996), and could be the reason why Saturn, as most giant exoplanets (see section 2.5), did not grow more massive than Jupiter: depending when gas accretion starts, there is not enough time to grow super giant planets.

This type of research is demanding, both in terms of computing capability and time, hence still open. In particular, non-isothermal simulations are necessary to better determine the structure of the CPD. Its temperature has strong implications for the composition of the solids available to form the satellites and for the chemical species in the gas. Novel and promising results on gas accretion by a giant planet are expected in the future.

### 2.3.4 Formation of Saturn, from consideration of the formation of moons and rings

For a comprehensive understanding of the formation and evolution of Saturn it is necessary also to gain an insight into the formation of Saturn's moons. Here we provide a brief discussion of this aspect of the Saturn system. About 60 satellites with confirmed orbits have been detected so far. Among them, 23 have quasi-circular orbits of radius smaller than 2 million kilometers in the plane of Saturn's equator: the so-called regular satellites. The others have eccentric, inclined (sometimes even retrograde), and larger orbits. They are called irregular satellites, and are supposedly captured. Hence, the irregular satellites do not inform much about Saturn's formation. In contrast, the rings and the regular satellites most likely formed together with Saturn in some way, and therefore provide constrains.

Titan, the largest moon of Saturn, dominates the population of satellites, being 60 times more massive than the second largest moon, Rhea. Titan's composition can provide further insight into the physico-chemical conditions prevailing in Saturn's CPD that must have played a crucial role in the make-up of Titan's building blocks. In situ measurements with the Huygens gas chromatograph mass spectrometer (Niemann et al., 2005, 2010) revealed for the first time that the bulk atmosphere of Titan is approximately 94% by volume nitrogen ($N_2$) and ~6% methane ($CH_4$). Methane may have originated on Titan, but direct external contribution is also possible. $N_2$, on the other hand, is almost certainly "secondary", i.e., instead of being delivered directly as $N_2$, it resulted from other nitrogen-bearing molecules originally captured in Titan's building blocks. Before nitrogen was actually detected on Titan by Voyager in 1980, Atreya et al. (1978) showed that the solar UV photolysis of ammonia ($NH_3$) could produce a substantial atmosphere of nitrogen on Titan in the past, which was eventually confirmed by Huygens in 2005. The dissociation of ammonia by impact shock heating has also been proposed (Jones and Lewis, 1987; McKay et al., 1988; Sekine et al., 2011; Ishimaru et al., 2011). While it seems like an attractive hypothesis, it faces insurmountable hurdles, including the removal of accompanying copious oxygen-bearing species and hydrogen, not found on Titan (see, e.g., Atreya et al., 2009). The fact that Titan's $N_2$ is not primordial but formed from ammonia has important implications for Saturn's CPD, considering possible scenarios of Titan's formation. Similarly, the origin of Titan's methane has a bearing on Saturn's CPD, so it is also discussed here briefly.

Two possibilities for the origin of Titan's methane have been proposed – production on Titan, or delivery to Titan. In the former case, methane was produced by hydro-geochemistry, i.e. water-rock reactions or serpentinization in the interior of Titan during its accretionary heating phase, when water was presumably in contact with the rocky core (Atreya et al., 2006; 2009). In this scenario, $H_2$ liberated in serpentinization reacts with primordial carbon in the form of CO, $CO_2$ or carbon grains in a metal-catalyzed Fischer-Tropsch process to produce methane. Mousis et al. (2009a) surmised that if water-rock reactions were responsible for Titan's methane, the D/H ratio in Titan's $CH_4$ (~$1.3\times10^{-4}$) should then reflect the value in Titan's water ice. As no measurements are available for D/H in Titan's water ice, they assumed that the D/H ratio measured in the water vapor plumes of Enceladus could serve as a proxy for the D/H in Titan's water. The Enceladus D/H value is $3\times10^{-4}$, which is more than twice the value in Titan's $CH_4$. This discrepancy led Mousis et al. (2009a) to propose that Titan's methane was trapped in its building blocks, which agglomerated from icy grains condensed in the protosolar nebula. In this scenario, Titan's methane would originate from ISM and its inferred D/H value would have resulted from the isotopic exchange with the nebula's hydrogen that occurred until it condenses and agglomerates by the building blocks of Titan. This conclusion was supported by the measurements of D/H in water in six Oort cloud comets available at that time, all of which have a value that is nearly identical to that measured in Enceladus' $H_2O$ plumes by the Cassini ion and neutral mass spectrometer, and corresponding to more than two times the value of D/H in Titan's methane. However, later observations of a Jupiter family comet Hartley 2 yield a D/H ratio of $1.56\times10^{-4}$ in water (Hartogh et al., 2011), which is similar to the value in Titan's $CH_4$ within the range of uncertainty for both objects. Another Jupiter family comet, 67P/Churyumov-Gerasimenko, on the other hand, yields D/H=$5.3\times10^{-4}$ in water (Altwegg et al., 2015), four times higher than in Titan's $CH_4$. Neither of these two comets has a D/H ratio in water similar to that assumed for Titan's water ice, the Enceladus value. Though the argument of methane trapping from the protosolar nebula appears to be weakened in view of these findings, it remains a plausible mechanism that needs to be validated by future observations, including, for example, direct D/H measurement in Titan's water, D/H in a large number of comets as well as laboratory studies and modeling.

Two main models have been proposed for the formation of large satellites of the gas giant planets, including Titan. In the first model, the satellite formation takes place in a dense and hot disk at the early stages of the gas giant planet formation (Prinn and Fegley, 1981, 1989; Lunine et al., 1989), while the other model uses a thin and cold disk to depict satellite formation (Canup and Ward, 2002). In the former model, the chemical composition of Titan's proto-atmosphere would have been primarily $CH_4$ and $NH_3$. These gases would have been produced, prior to planetesimal condensation, from CO and $N_2$ initially present in the dense, hot and chemically active CPD (Prinn and Fegley, 1989; Sekine et al., 2005). This scenario may be ruled out for the bulk of Titan's nitrogen on the basis of the nitrogen isotope ratio, with the caveat about atmospheric escape, as discussed below. In Titan's atmosphere, $^{14}N/^{15}N$ = 167.7 (Niemann et al., 2010), which is much less than, not similar to, the value in Saturn (>357; Table 2.3, Section 2.2.1), which implies that the ammonia accreted by Titan did not originate from the protosolar nebula. On the other hand, the present nitrogen isotope ratio depends on the evolutionary history and the processes of escape of nitrogen from Titan early on and in the past 4.5

Gyr, which are very poorly constrained. Even on Earth and Venus that have atmospheres as dense or even denser than Titan, escape processes have shaped their present atmospheric isotope ratios. Additional modeling and observations are needed to resolve the issue of evolution of Titan's nitrogen isotope ratio over time.

The other model suggests that icy planetesimals were actively supplied into the CPD from Saturn's feeding zone in the solar nebula (Canup and Ward, 2002; Alibert and Mousis, 2007). In this scenario, the chemical composition of Titan's proto-atmosphere would derive from that of CO- and $N_2$-rich icy planetesimals formed at low temperature (~20 K) in the protosolar nebula (see Section 2.4.1 for details concerning the composition of the protosolar nebula). However, similar to the previous case, this scenario is found inconsistent with the low $^{14}N/^{15}N$ ratio measured in Titan's nitrogen, but with the caveat of nitrogen escape mentioned above. In order to solve these discrepancies, Mousis et al. (2009b) proposed that Titan was formed from icy planetesimals initially produced in the solar nebula and that were partially devolatilized during their migration within Saturn's CPD. By doing so, Titan's building blocks preserved the ammonia and methane they acquired from the protosolar nebula and released most of the carbon monoxide and nitrogen prior to satellite formation. However, as discussed above, production of methane on Titan by serpentinization, rather than direct delivery of $CH_4$, is a very attractive mechanism.

Considering the lack of full complement of relevant observational constraints for Titan, Saturn, and the comets clear discrimination between the two scenarios is not possible at this time. It is also plausible that both mechanisms could have played a role to a varying degree. Nevertheless, above considerations about Titan's composition still allow us to place important constraints on the thermodynamic state of Saturn's CPD at the time of formation of its largest satellites. In view of the low $^{14}N/^{15}N$ ratio measured in Titan's atmosphere, it seems that the CPD may not have been warm and dense enough to allow in situ condensation of its building blocks, but available observational constraints are insufficient to make a firm statement, as discussed above. On the other hand, a temperature-density gradient did probably exist throughout Saturn's CPD, not important enough to allow the vaporization of water ice at Titan's formation zone, but probably sufficient to explain why Titan's primordial nitrogen reservoir is $NH_3$

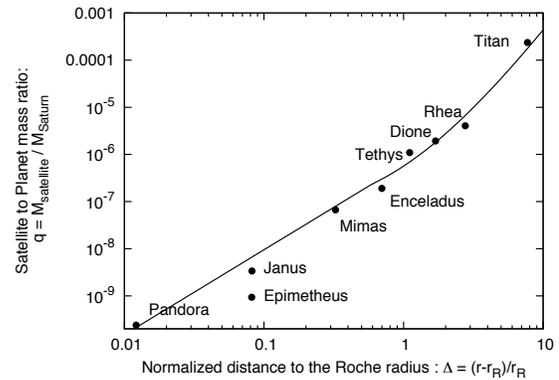

**Figure 2.6.** Satellite to planet mass ratio q as a function of the normalized distance to the Roche radius, located at the outer edge of the rings, 140,000 km away from Saturn's center. The line corresponds to the model described in Section 2.3.4.

and not $N_2$ as is the case for Saturn (Mandt et al., 2014).

The above two models focus on the dominant body only (Titan) and somehow disregard the system of regular satellites of Saturn as a whole. However, a recent scenario for the formation of the regular satellites interior to Titan provides constrains on Saturn's history and internal structure. The rings spread, and spread faster when they are more massive, so that after about 4 Gyr of evolution, they should have roughly the present mass, whatever their initial mass (Salmon et al., 2010). Hence, it is possible that they originally were thousands times more massive than now (e.g. Canup 2010). As the rings spread beyond the Roche radius, they agglomerate into moonlets, which migrate outwards due to their interaction with the rings (Charnoz et al., 2010). Numerical simulations show that this process can generate the 10 regular satellites inside Titan, and even explain the irregular silicate cores of the 5 largest ones (Charnoz et al., 2011). Crida and Charnoz (2012) solved analytically the equations governing the formation and migration of satellites from the spreading of rings beyond the Roche limit and found that the mass-distance relation in a system of satellites formed in this manner must follow a particular power law, which is represented in Figure 2.6. The agreement with the observed distribution supports this model. Even Titan lies on the theoretical line, which could be a coincidence, or not. Iapetus doesn't fit in this model and is not shown in the figure (it would be further on the middle right), but Iapetus is thought to have formed concurrently with Saturn in the circum-planetary disk (Castillo-Rogez et al., 2009).

It is possible that the regular satellites inside Titan formed after Titan and Saturn, from the spreading of initially massive rings. But for this to happen within the age of the solar system, strong tidal dissipation is needed inside Saturn. Tidal dissipation is characterized by a dimensionless factor, generally denoted as Q; the lower the value of Q, the stronger the dissipation. Charnoz et al. (2011) found that with Q of the order of 1700, as argued by Lainey et al. (2012, 2015), the formation of Saturn's satellite system takes about 3.5 Gyr (present satellite crater density record does not provide unambiguous evidence either in favor of or against this timescale). The standard value of Saturn's Q used to be an order of magnitude larger (corresponding to ten times less dissipation), but one should consider that this high value was supported by the constraint that the satellites were supposed to have hardly moved since the formation of the solar system, which may be incorrect. In contrast, the Lainey et al. value is based on observations, and consistent with this new model for satellite formation. In the end, models of the formation of the satellites allow us to place constraints on the efficiency of the dissipation inside Saturn, hence on its interior. Remus et al. (2012) show that low values of Q (high dissipation rates) are possible in the framework of a model in which tidal dissipation occurs at the interface between the core and the envelope as a result of different anelastic deformations. Values as low as $10^3$ require a specific range of values of the shear modulus and viscous modulus in the core. Unfortunately these two quantities are almost unknown given the uncertainties on the size, composition and physical state of the core.

## 2.4. Saturn's formation: chemical point of view

Just as the hydrodynamical scenario discussed in the previous section provides insight into Saturn's origin, chemical evolution of the protosolar disk and the manner in which volatiles are sequestered in grains or planetesimals together with their nature and delivery to Saturn provide valuable constraints to the models of Saturn's formation and evolution. This section elaborates on these processes.

### 2.4.1 Chemical evolution of the protosolar disk

Formation scenarios of the protosolar nebula invoke two reservoirs of ices, namely an inner and an outer reservoir, which took part in the production of icy planetesimals. The first reservoir contains ices (mostly water ice) originating from the ISM, which were initially vaporized due to their proximity to the Sun. With time, the decrease of temperature and pressure conditions allowed the water in this reservoir to condense at ~150 K in the form of microscopic crystalline ice on the surface of pre-existing refractory grains (Kouchi et al., 1994). The other reservoir, located at larger heliocentric distances, is composed of ices originating from the ISM that were preserved when entering into the disk. In this reservoir, water ice was essentially in the amorphous form and the other volatiles remained trapped in the amorphous matrix (Notesco and Bar-Nun, 2005). The exact localization of the boundary between these two reservoirs, corresponding to the initial location of the so-called "iceline", depends on the assumed thermal structure of the disk, which is still poorly constrained. Optically thin disks such as the steady nebula model of Hayashi (1981) predict that the water iceline is located just beyond the Main Belt (also known as the Asteroid Belt, which is located between the orbits of Mars and Jupiter, or 2.2-3.2 AU). On the other hand, optically thick models of the protosolar nebula suggest that the water iceline might have been initially up to ~30 AU from the Sun (Chick and Cassen, 1997).

The O-, C- and N-bearing ices delivered from ISM to the protosolar nebula are expected to be essentially constituted from $H_2O$, CO, $CO_2$, $CH_3OH$, $CH_4$, $N_2$ and $NH_3$, with $H_2O$, CO, $CO_2$ and $N_2$ being the most abundant molecules in decreasing order (Öberg et al., 2011a; Gibb et al., 2004). $H_2O$ ice is expected to be dominant because i) of its high abundance (due to the cosmic abundances of H and O) and ii) it is by far the first volatile to condense as the temperature decreases in the nebula.

Regardless of the considered volatile reservoir in the protosolar nebula, dust and ice particles coagulated until they reached cm-sized pebbles. Once formed, these pebbles agglomerated into large planetesimals (10-1000 km) by streaming instabilities (Youdin and Goodman, 2005; Johansen and Youdin, 2007; Johansen et al., 2009) and formed the cores of the giants on timescales that were sufficiently short to allow in situ formation of these planets prior to their migration in the protosolar nebula (Lambrechts et al., 2014). Pebbles and planetesimals formed in the outer reservoir should have coagulated from pristine amorphous ice originating from ISM. In contrast, pebbles and planetesimals formed during the cooling of the inner reservoir coagulated from a mixture of microscopic icy grains made of pure condensates, stoichiometric hydrates (such as NH$_3$-H$_2$O) and clathrates, whose proportions depended on the availability of water ice and the temperature to which the disk had cooled down.

### 2.4.2 Delivery of volatiles to Saturn via the accretion of planetesimals

Several hypotheses relating the thermodynamic evolution of the protosolar nebula to the formation conditions of the giant planets have been developed in order to interpret their observed volatile enrichments. In particular, the volatile enrichments observed in the giant planets can be explained by the accretion of icy planetesimals and their vaporization in the envelopes at the time of their growth from nebular gas. The two main scenarios proposed in the literature, each based on the hypothesis that the giant planets accreted planetesimals originating from one of the two abovementioned reservoirs of ices, are discussed below.

*Delivery of amorphous ices to Saturn*
Owen et al. (1999) proposed a cold icy planetesimal model, according to which the volatile enrichments observed by the Galileo probe in Jupiter result from the accretion of planetesimals agglomerated from amorphous ice at temperatures below approximately 30 K (such low temperatures are needed to trap N$_2$ and Ar; Owen et al., 1999). Owen et al. postulated that either Jupiter was formed at large heliocentric distances of 40-50 AU where the cold temperature favored the preservation of amorphous ice in the disk and then migrated to its current location, or the protosolar nebula was much cooler at the current location of Jupiter (~5 AU) than what is predicted by current turbulent accretion disk models. In either case, the icy material originated from the protosolar cloud and survived the formation of the protosolar nebula. If correct, this scenario predicts that the volatile enrichments should be uniform (also, Owen and Encrenaz, 2006) since volatiles are not fractionated when trapped in amorphous ice. However, as discussed in Section 2.2.1, current analysis of the Galileo Probe data shows that the enrichment of the observed heavy elements spans a range of 2 to 6 times the current solar elemental abundances. For Saturn, key data to assess the validity of the icy planetesimal model or another model are presently lacking. Noble gases are not measured. Amongst non noble gases, the only heavy element with a robust value is carbon, with C/H approximately 9× solar (Table 2.1). Sulfur enrichment is similar to carbon, but the result is tentative (Section 2.2.1). NH$_3$ is a good measurement, but currently provides N/H only in the 1-3 bar region. As discussed earlier (Section 2.2.1), it is far from certain that the N/H value in the deep well-mixed atmosphere of Saturn is going to be similar; it could be greater. If one assumes that the N/H in Saturn's deep atmosphere is unchanged from the value at 3 bars, then the C-enrichment is greater than N-enrichment by a factor of 3, not same, which would argue against the cold icy planetesimal model as presented in Owen et al. (1999). On the other hand, C/S would favor it, if the H$_2$S result were confirmed by future observations.

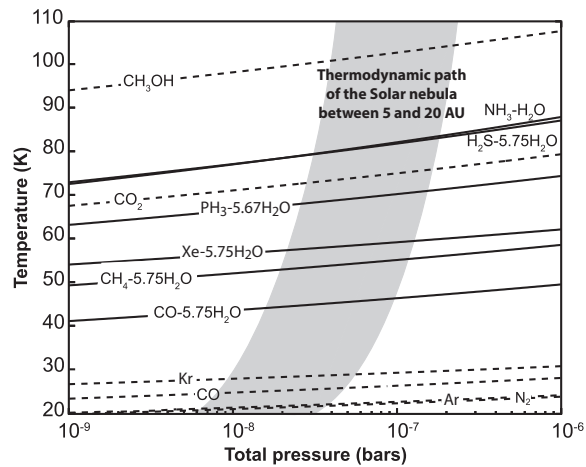

**Figure 2.7.** Formation conditions of icy planetesimals in the solar nebula. Equilibrium curves of hydrate (NH$_3$–H$_2$O; solid line), clathrates (X-5.75H$_2$O or X-5.67H$_2$O; solid lines), and pure condensates (dashed lines) are superimposed with the thermodynamic path of the solar nebula in the 5-20 AU range, assuming full clathration efficiency. Elemental abundances are solar, with molecular ratios specified in Mousis et al. (2012). Species remain in the gas phase above the equilibrium curves. Below, they are trapped as clathrates or simply condense. The clathration process stops when no more crystalline water ice is available to trap the volatile species.

*Delivery of crystalline ices to Saturn*
An alternative interpretation of the volatile enrichments measured in Jupiter is based on the hypothesis that most of volatiles were trapped in clathrates in the giant planet's feeding zone (Gautier et al., 2001; Alibert et al., 2005a; Alibert et al., 2005b). These authors assumed that Jupiter's building blocks formed in the inner zone of the protosolar nebula, in which the gas phase has been enriched at early epochs by the vaporization of amorphous ice entering from the Interstellar Medium (ISM). During the cooling of this region of the disk, water vapor crystallized and trapped the volatiles in the form of clathrates or stoichiometric hydrates in the 40-90 K range instead of condensing at lower temperatures. These ices then agglomerated and formed the solids that were ultimately accreted in the envelope of the growing Jupiter. These scenarios, which assume the full (100%) clathration of volatiles, are based on the hypothesis that the amount of available crystalline water ice was large enough ($H_2O/H_2 > 2\times(O/H)_{protosolar}$) to trap the other volatiles in the feeding zone of Jupiter. Later studies have shown that it is also possible to explain the volatile enrichments in Jupiter via the accretion and the vaporization in its envelope of icy planetesimals made from a mixture of clathrates and pure condensates (Mousis et al., 2009c; Mousis et al., 2012), assuming a full protosolar composition for the gas phase of the disk and provided that the disk's temperature decreased down to ~20 K at their formation location. Figure 2.7 represents a clathration/condensation sequence of volatiles that has been used by Mousis et al. (2012) to interpret the volatile enrichments in Jupiter.

In the case of Saturn, if only C enrichment is considered (Section 2.2.1), it is easily explained via the delivery of planetesimals formed at similar low temperatures as those accreted by Jupiter. When considering both C and N enrichments measured in Saturn with C/N=3 provided that N/H in the bulk atmosphere is the same as in 1-3 bar region, which is far from certain (Section 2.2.1), the scenario of full volatile clathration may not hold anymore because it would result in a uniform enrichment of these two species (Mousis et al., 2006). It has thus been argued that Saturn might have formed at a higher temperature than those required for the formation of CO and $N_2$ clathrates in the protosolar nebula (Hersant et al., 2008). However, this scenario does not match the high $^{14}N/^{15}N$ ratio recently estimated for Saturn (>357, Table 2.3; Section 2.2.2), since it predicts a value intermediate between the value for Jupiter (434) and the Earth (272). As discussed above, much of the critical heavy element abundance data for Saturn are missing to fully evaluate the validity of the clathrate model, and to some extent they are missing also for Jupiter. With 100% efficiency of clathration, models predict approximately 15× solar O/H at Jupiter (e.g. Gautier et al., 2001, using current solar O/H of Asplund et al., 2009), whereas the icy planetesimal model predicts it to be four times less (Owen et al., 1999). Water is critical for discriminating between various formation scenarios. Little laboratory data are presently available for the relatively low pressure conditions of the solar/protoplanetary nebula. In summary, both the cold icy planetesimal model and the clathrate model have their strengths and weaknesses, and to discriminate between them requires new sets of data, particularly for Saturn (see chapter of Baines et al. for additional details).

### 2.4.3 Role of photoevaporation of the protosolar disk in determining present-day composition

The atmospheres of the giant planets result from the accretion of both gaseous and solid material by planetary cores. The clathrate scenario implicitly assumes that all species other than hydrogen and helium were delivered with the solids. However, processes affecting the protosolar disk itself may also play an important role in determining the final from the central star (Gorti et al., 2009) and by ambient FUV irradiation from other stars in the cluster (Adams et al., 2004). This evaporation takes place in the disk atmosphere, a region in which the temperature gradient is strongly negative (Chiang and Goldreich, 1997). This would prevent a convective transport of species in the mid-plane regions and therefore, Guillot and Hueso (2006) conjecture that hydrogen and helium would preferentially evaporate. This would lead to a progressive homogeneous enrichment of the disk in condensing species. If formed late, giant planets would incorporate gas that is heavy-element rich, and in particular it would be rich in species such as Ar, Kr and Xe. This theory explains the enrichment in noble gases in Jupiter's atmosphere measured by the Galileo probe and predicts a similar, homogeneous enrichment in Saturn atmospheric compositions. A plausible scenario proposed by Guillot and Hueso (2006) to explain the homogeneous enrichment of noble gases in Jupiter is illustrated in Figure 2.8. It is based on the fact that protoplanetary disks can extend to hundreds of AU and that their outer parts are generally very cold (e.g. Dartois et al., 2003). Temperatures of 10-30 K in the outer disks allow the direct condensation of most noble gases onto small grains (e.g. Owen et al., 1999). These grains will grow, settle towards the disk mid-plane and migrate inward (e.g. Adachi et al., 1976; Weidenschilling, 1984; Dubrulle et al., 1995). In parallel, the gas disk is being accreted by the central star and photoevaporated both due to direct irradiation (i.e. with solar Kr/Ar and Xe/Ar ratios). It cannot make predictions on elements that are

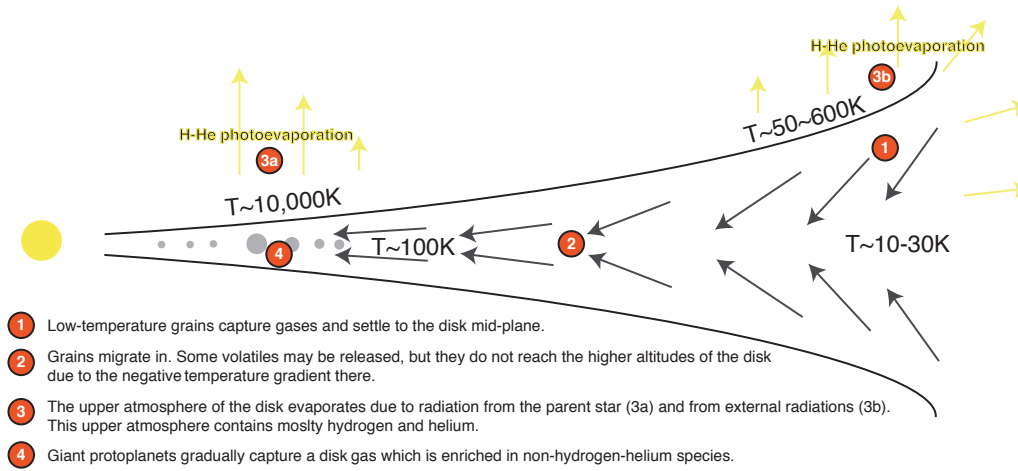

1. Low-temperature grains capture gases and settle to the disk mid-plane.
2. Grains migrate in. Some volatiles may be released, but they do not reach the higher altitudes of the disk due to the negative temperature gradient there.
3. The upper atmosphere of the disk evaporates due to radiation from the parent star (3a) and from external radiations (3b). This upper atmosphere contains moslty hydrogen and helium.
4. Giant protoplanets gradually capture a disk gas which is enriched in non-hydrogen-helium species.

**Figure 2.8:** Sketch illustrating the scenario proposed by Guillot and Hueso (2006) to explain a homogeneous enrichment of noble gases in the envelopes of giant planets. A gaseous protosolar disk is shown edge-on. Forming protoplanets are shown by grey circles. Black arrows represent the dynamical evolution of grains and noble gases. Yellow arrows correspond to photoevaporation of gas from the disk due to both internal and external UV irradiation. The 4 circles correspond to the important evolution steps from the condensation of noble gases into cold grains in the outer disk to their incorporation in the envelopes of growing giant planets.

delivered into giant planets with the solids and for which the story may be more complicated, as illustrated in the previous section.

## 2.5 Extrasolar giant planets context

The discoveries of numerous extrasolar planets in recent years are now allowing us to place the solar system planets in a cosmic context. Over 3000 confirmed exoplanets[1] are known as of June 2016, of which over 500 are giant planets larger (in mass and/or size) than Saturn. The majority of these planets have been detected either through doppler spectroscopy of their host stars, i.e. the 'radial velocity' (RV) method, or by observing transits of the planets in front of their host stars (i.e. the 'transit' method), while a few tens of the planets have been detected via direct imaging. The observational sensitivities of the various exoplanet detection methods have precluded conclusive detections of exact analogues of Saturn and Jupiter in exoplanetary systems. Currently, the RV and transit methods, which together allow measurements of masses and radii of exoplanets are preferentially sensitive to planets at short orbital separations. Giant exoplanets with both masses and radii measured are known at orbital separations of ~0.01 AU – 0.5 AU, implying

---

[1] Extrasolar Planets Encyclopedia (exoplanet.eu) list of currently known exoplanets and their properties

equilibrium temperatures over ~1000 K. On the other hand, while the direct imaging method is more sensitive to planets with large orbital separations (≳10 AU), current instruments are only sensitive to young, and hence also hot, giant planets whose large fluxes make them detectable. These diverse giant exoplanets form a starting point for placing the properties of solar system giant planets in perspective. In this section, we review our current understanding of the interiors, atmospheres, and formation conditions of extrasolar giants planets and their analogies with Saturn and Jupiter in the solar system.

### 2.5.1 Interiors of giant exoplanets

Constraints on the interior compositions of giant exoplanets are based primarily on their masses and radii, which are both known for about 200 transiting exoplanets with masses and radii greater than those of Saturn (0.30 $M_J$ and 0.84 $R_J$), generally referred to as 'hot Jupiters'. These planets have revealed an extreme diversity in their bulk parameters with masses of 0.3 – 20 $M_J$, radii of 0.84 – 2 $R_J$, and temperatures of ~1000-3000 K. The masses and radii of these planets are consistent with gaseous interiors dominated by $H_2$ and He, similar to Saturn and Jupiter in the solar system. However, the diversity in masses and radii also imply a wide range of possible core masses, ranging from no core to ~200 $M_E$ for the heaviest planets (e.g. Guillot et al., 2006; Baraffe et al., 2008), while upper-limits on the core mass in

Saturn and Jupiter are estimated to be ~10-20 $M_E$ as discussed in Section 2.2.3. Some studies have also suggested a possible positive correlation between the planetary core mass in hot Jupiters and the metallicity of the host star (e.g. Guillot et al., 2006). However, constraints on the core masses in giant exoplanets and on their internal structures in general are confounded by several complexities (see e.g. review by Baraffe et al., 2014).

Unlike the solar system giants, a large fraction of hot Jupiters have radii that are significantly larger than those predicted by internal structure models of giant planets even assuming no solid core; the presence of a core contracts the planet. The largest planet known, WASP-17b, has a radius of ~2 $R_J$ but a mass of only 0.5 $M_J$ (Triaud et al. 2010). Such bloated giant planets have no analogies in the solar system, and cannot be explained by canonical interior models of Saturn and Jupiter. Several mechanisms have been proposed to address the problem which bear on the key factors in which hot Jupiters differ from solar system giant planets. Hot Jupiters receive stellar irradiations that are 3-4 orders of magnitude higher than the insolation at Jupiter, and because of their close proximity to their host stars they are also subject to strong tidal and magnetic interactions. The various proposed mechanisms to explain inflated giants, though still actively debated, broadly include (a) deposition of incident energy deep in the planetary atmosphere causing an extra energy source in the planetary interior and slowing down the thermal evolution, i.e. cooling, of the planet (Guillot and Showman, 2002; Batygin and Stevenson, 2010; Youdin and Mitchell, 2010) (b) intrinsic heating caused by tidal dissipation in the planetary interior due to an eccentric close-in orbit that is being tidally circularized (e.g. Bodenheimer et al., 2001; Leconte et al., 2010), (c) strong atmospheric opacity that inhibits the emergent flux thereby delaying the cooling, and hence contraction, of the planet during its evolution (Burrows et al., 2007). However, none of these mechanisms conclusively explains the radii distributions in all the hot Jupiters currently known (see e.g. Spiegel and Burrows, 2013; Baraffe et al., 2014). Consequently, even though hundreds of giant exoplanets are known with similar masses and sizes as solar system giants, their interior structures and compositions are likely extremely diverse albeit currently underconstrained.

### 2.5.2 Atmospheres of giant exoplanets

Remarkable progress has been made in the past decade in spectroscopic observations of exoplanetary atmospheres, primarily of hot gas giants that are most accessible to current instruments (see e.g. review by Madhusudhan et al., 2014b). Currently observable gas giant atmospheres fall into two distinct categories, (a) highly irradiated giant planets ('hot Jupiters') in very close orbits (as close as 0.01 AU), and (b) young and self-luminous directly-imaged planets at wide orbital separations (beyond ~10 AU). The effective temperatures of either class of planets are in the range of ~1000-3000 K, which are an order of magnitude hotter than those of solar system giant planets (~100-200 K). Since the radiation field is intricately linked to the physicochemical characteristics of the atmospheres, the atmospheric temperature structure, chemistry, and dynamics in these giant exoplanets can be markedly different from those of Saturn and Jupiter in the solar system, even if the masses, radii, and bulk elemental abundances turn out to be similar. Here, we review current understanding of giant exoplanetary atmospheres vis-a-vis our understanding about the atmospheres of Saturn and Jupiter.

*Atmospheric observations*
Spectra of exoplanetary atmospheres are inherently disk-integrated, unlike spectra of solar system giant planets, which can be spatially resolved over the planetary disk. Observations of exoplanetary spectra have been obtained using three key methods. Firstly, the atmospheres of close-in hot Jupiters have been observed primarily through transit spectroscopy, obtained during the planet's 'transit' in front of the host star or 'occultation' behind the star. While a transit (or transmission) spectrum probes the atmosphere of the day-night terminator region of the planet, the occultation (or emission) spectrum probes the dayside atmosphere of the planet. Spectra of transiting hot Jupiters have been observed both from space, using the Hubble and Spitzer space telescopes, as well as from ground-based facilities. While Spitzer and ground-based facilities have typically provided photometric observations of transiting exoplanets in the near-infrared (e.g. Charbonneau et al., 2008; Croll et al., 2011), the Hubble telescope has been instrumental in obtaining spectra across multiple spectral regimes from the ultraviolet to near-infrared for a few planets (Vidal-Madjar et al., 2003; Sing et al., 2011; Deming et al., 2013). These state-of-the-art observations have provided both the high precision and a long spectral baseline required to constrain the atmospheric properties of several transiting hot Jupiters. Secondly, it has also been possible to detect molecules in the atmospheres of a few transiting and non-transiting close-in hot Jupiters using very high resolution ($R \sim 10^5$) infrared doppler spectroscopy using large ground-based telescopes (Snellen et al., 2010). Thirdly, ground-based spectroscopy of directly imaged planets has led to both photometry and high resolution spectra of thermal emission from several young self-luminous planets in the near infrared (e.g. Marois et al., 2010; Konopacky et al., 2013; Janson et al., 2013).

*Atmospheric chemistry*
Chemical compositions of hot giant exoplanets are expected to be markedly different from those of solar system giant planets, even if the bulk elemental abundances may be identical. The bulk molecular composition of the atmospheres of Saturn and Jupiter is generally consistent with expectations for low temperature (~100-200 K) H-rich atmospheres, i.e. dominated by methane, ammonia, and higher-order hydrocarbons (Section 2.2.1). $H_2O$ is expected to be the dominant O carrier but its abundance is presently undetermined in both Saturn and Jupiter because of their low temperatures as discussed in Section 2.2.1. On the other hand, $H_2O$ is more observable in the high-temperature atmospheres of giant exoplanets. However, the expected molecular composition depends strongly not only on the atmospheric temperatures but also on the elemental abundance ratios, particularly the overall metallicity and the C/O ratio (Madhusudhan, 2012; Moses et al., 2013). Assuming solar abundances (i.e. C/O = 0.5), in the 1000-3000 K temperature range of hot giant exoplanets, $H_2O$ is expected to be the dominant carrier of O in the observable atmosphere, CO is expected to be the dominant C carrier above ~1300 K while at lower temperatures $CH_4$ and $NH_3$ are expected to be abundant, along with trace quantities of $CO_2$ (Lodders and Fegley, 2002; Madhusudhan, 2012). Other species expected in hot Jupiters include Na, K, TiO, and VO (Seager et al., 2000; Hubeny et al., 2003; Madhusudhan, 2012), which are not found in solar system gas giants because of their low temperatures. The chemistry can be even more drastic for super-solar abundance ratios, e.g. C/O = 1 in which case $H_2O$ can be substantially underabundant and carbon-rich species overabundant even in very high temperature atmospheres (Madhusudhan, 2012; Moses et al., 2013). Therefore, molecular abundances in hot Jupiters serve as key indicators of their elemental abundance ratios, such as the C/O ratio.

Chemical species have been detected in several giant exoplanetary atmospheres using all three observational methods discussed above. Recently, $H_2O$ has been detected at high statistical significance in the atmospheres of several transiting hot Jupiters using the HST WFC3 spectrograph in the near-infrared (1.1-1.7 μm), e.g. in HD 209458b, HD 189733b, WASP-43b, and WASP-17b (Deming et al., 2013; Mandell et al., 2013; Kreidberg et al., 2014; McCullough et al., 2014). Additionally, transmission spectroscopy in the visible has been used to detect several atomic species in hot Jupiter atmospheres, e.g. Na and K (Redfield et al., 2008; Sing et al., 2011).

More recently, CO and $H_2O$ have been detected in some transiting as well as non-transiting hot Jupiters using ground-based high-resolution infrared doppler spectroscopy (e.g. Brogi et al., 2012; Birkby et al., 2013). On the other hand, $H_2O$, CO, and $CH_4$ have also been detected robustly in the atmospheres of directly imaged young giant exoplanets using high resolution ground-based spectroscopy (e.g. Janson et al., 2013; Konopacky et al., 2013). The recent detection of a methane-rich giant exoplanet 51 Eri b roughly twice the mass of Jupiter (Macintosh et al., 2015) represents the closest, albeit young (~20 Myr), analogue to solar-system giant planets. 51 Eri b orbits a Sun-like star 51 Eridiani at a Saturn-like orbital separation (13 AU), and like Saturn and Jupiter contains $CH_4$ as the dominant C-bearing molecule in its atmosphere.

In addition to molecular detections, recent observations are beginning to place notable statistical constraints on the molecular abundances in giant exoplanetary atmospheres suggesting likely diverse elemental compositions. On one hand, some of the highest precision HST WFC3 near-infrared spectra of transiting hot Jupiters are revealing significantly weaker $H_2O$ features than expected for solar-composition atmospheres. For example, the thermal emission spectrum of the hot Jupiter WASP-12b suggest 100× sub-solar $H_2O$, and a C/O ≥ 1, in its dayside atmosphere (Madhusudhan, 2012; Stevenson et al., 2014), and that of WASP-33b suggests ~5-10× sub-solar $H_2O$ but with C/O < 1 (Haynes et al., 2015). Similarly, high-precision transmission spectra of the day-night terminator regions of hot Jupiters HD 189733b and HD 209458b suggest $H_2O$ abundances as low as 100× sub-solar, assuming cloud-free models (Deming et al., 2013; Madhusudhan et al., 2014b). It is possible that the presence of high-temperature silicate clouds/hazes as discussed below could be masking the spectral features in some transmission spectra. On the other hand, thermal emission and transmission spectra of the hot Jupiter WASP-43b (semimajor axis 0.01526 AU, orbital period 0.81 days, planetary mass 2 $M_J$, host star mass 0.717 $M_\odot$ and $T_\odot$ 4520 K) reveal $H_2O$ abundances in the range 0.4-3.5× solar at 1σ confidence level and an upper limit of 20× solar at 3σ confidence level (Kreidberg et al., 2014). Thus, there is a real possibility of super-solar O/H in at least some extrasolar giant planets. Only when $H_2O$ is measured in well-mixed atmospheres of Saturn and Jupiter, direct comparison with O/H in exoplanets will be possible. Meanwhile, consistency between the super-solar O/H in WASP-43b and super-solar C/H in all solar-system giant planets and super-solar Ar, Kr, Xe, N and S in Jupiter seems to indicate similar formation processes of at least some hot Jupiters and Jupiter and Saturn in the solar system, but much further work is needed to be confident.

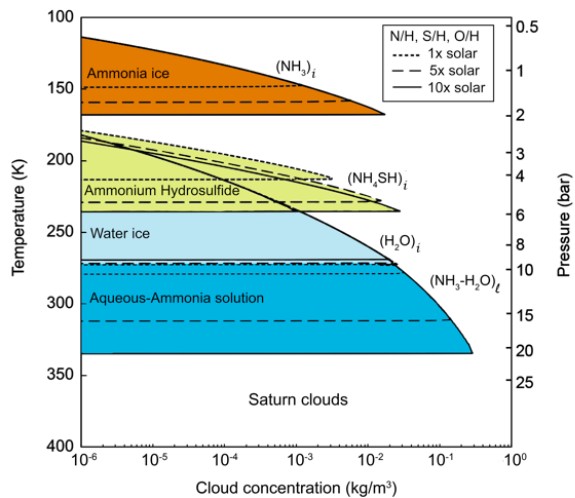

**Figure 2.9.** Equilibrium cloud condensation model of Saturn, assuming uniform enrichment of 1× solar, 5× solar and 10× solar abundances for each of the condensible volatiles, $NH_3$, $H_2S$ and $H_2O$, hence of the elemental ratios N/H, S/H and O/H, respectively. Calculatiosn are based on current solar elemental abundances (Asplund et al., 2009) from Table 1. [Note that $H_2S$ does not directly condense, but $NH_4SH$ produced by the vapor-phase reaction between $H_2S$ and $NH_3$, does in solid form [$(NH_4SH)_i$]. Water can condense as ice $(H_2O)_i$, and liquid of water-ammonia solution [$(NH_3$-$H_2O)_l$, colloquially, windex cloud] for large enrichment of $H_2O$. The cloud concentrations represent *upper limits*. Precipitation and dynamics would almost certainly deplete cloud densities by up to several orders of magnitude, as in the water clouds in the Earth's troposphere. The cloud bases (or, lifting condensation levels) are robust, however. More realistic cloud densities are formulated in Wong et al. (2015).

*Clouds and hazes*
Clouds are ubiquitous in Saturn and Jupiter, but with quite different chemical compositions (Sections 2.2.1 and 2.6) from those expected in giant exoplanets. While clouds in Saturn and Jupiter are presumably made of low-temperature (150-300 K) condensates of ammonia, hydrogen sulfide (combined with ammonia) and water, as discussed in Section 2.2.1 and illustrated in Figure 2.9, those in hot giant exoplanets (at 1000-3000 K) are expected to be composed of refractory compounds, such as silicates, alkali chlorides, Fe, etc. (Sudarsky et al., 2003). To date there is no spectral signature of a cloud-forming condensate in an exoplanetary atmosphere. Instead, the inferences of clouds in these atmospheres are derived from non-detections of expected atomic or molecular features (i.e. due to possible obscuration from clouds) or from modulations in the planetary spectrum indicative of particulate scattering, e.g. a steeper slope rising blue-ward than expected from pure gaseous Rayleigh scattering). For example, non-detections of strong Na and K absorption in the visible wavelengths (at 589 nm and 770 nm, respectively) along with a steep power-law spectrum have been suggested as indicative of haze in the hot Jupiter HD 189733b (Sing et al., 2011; Pont et al., 2013). The presence of clouds has also been inferred from observations of visible reflected light and phase curves of hot Jupiters (Demory et al., 2013; Evans et al., 2013; Barstow et al., 2014). On the other hand, several studies have used near-infrared observations of thermal emission from directly imaged giant planets, such as HR 8799b,c,d,e to suggest the presence of thick clouds in their atmospheres (e.g. Marois et al., 2010; Currie et al., 2011; Marley et al., 2012). Although hazes and clouds are not expected to be made up of $H_2O$ in hot Jupiters due to their atmospheric temperatures that are high enough to vaporize water, they could still remove O from gas phase in the form of condensed silicates thereby decreasing the amount of O available for $H_2O$. They could also provide adsorption/sequestration sites for water vapor (and other volatiles) that may result in the removal of $H_2O$ by heterogeneous chemistry or surface processes, depending on the nature of hazes and temperatures.

*Temperature profiles and stratospheres*
Accurate determination of the atmospheric temperature profiles is important to constrain various thermal processes in exoplanetary atmospheres, and also because the temperature gradient is degenerate with chemical composition in their contributions to an emission spectrum (Madhusudhan and Seager, 2010). One of the long-standing conundrums in the field concerns the possibility of temperature inversions (or 'stratospheres') in exoplanetary atmospheres, i.e. temperature increasing with altitude in the atmosphere, as opposed to a monotonically decreasing temperature profile, which would be expected for an isolated body. The Earth and larger planets in the solar system all have thermal inversions; on Earth it is due to ozone, in giant planets it is due to hydrocarbon haze. Early theoretical studies (Hubeny et al., 2003; Fortney et al., 2008) predicted that atmospheres of hot Jupiters could also host thermal inversions, but due to very different sources than those in Jupiter or Saturn, namely from gaseous TiO and VO which can survive at high temperatures. Current observations suggest that some hot Jupiters show 'tentative' evidence for thermal inversions whereas others do not (e.g. Stevenson et al., 2014; Haynes et al. 2015). Various processes have been proposed to explain possible trends, e.g. correlations with stellar irradiation (Fortney et al., 2008), TiO condensation (Spiegel et al., 2009), stellar-activity (Knutson et al., 2010), C/O ratios (Madhusudhan, 2012), and thermo-resistive instability (Menou, 2012).

Overall, there are presently no conclusive constraints on the presence of thermal inversions in exoplanetary atmospheres or on any inversion causing absorbers. High-resolution spectra from future facilities would be required to make robust detections of thermal inversions. Although detailed data on thermal structure of the atmospheres of Saturn and Jupiter exists as a result of spacecraft remote sensing and entry probe (at Jupiter) measurements, thermal structure of cold gas planets is not a suitable guide for what to expect in hot Jupiters whose structure is controlled by extreme stellar forcing. However, with appropriate modifications radiative transfer models used for interpreting temperature observations of cold, clear, cloudy or hazy gas planets are to some degree applicable to hot Jupiters (Lee et al. 2012).

### 2.5.3 Formation of giant exoplanets

The large population of giant exoplanets provides a diverse sample to test theories of formation of giant planets in the solar system. As discussed in Section 2.1, two main formation mechanisms have been proposed to explain the formation of Saturn and Jupiter: core accretion (CA) and gravitational instability (GI), with clear preference for CA. Various efforts have been made to identify if either of these formation mechanisms could explain the formation of giant exoplanets based on their observed orbital parameters and chemical compositions.

*Dynamical constraints*
The diverse orbital parameters of different classes of giant exoplanets (close-in versus distant) constrain the different possible formation mechanisms. In the CA model (Pollack et al., 1996), the planetary embryos start out as ~10 $M_E$ cores in the protoplanetary disk that subsequently undergo runaway accretion of a large volume of gas and planetesimals to form a massive gaseous envelope. On the other hand, a GI in a young disk can cause rapid collapse of a large volume of ambient gas and solids to form a giant planet (Boss et al. 2000). Both scenarios occur in planet-forming disks, but at different orbital separations. While CA is favored closer to the snowline (within ~2-10 AU) because cores take too long to form at larger distances and only reach large masses after the disk has dispersed, GI is favored at larger distances (≳10 AU) where the disk can cool sufficiently on orbital timescales to fragment. In this regard, GI may be the favored mechanism for the formation of distant gas giant exoplanets detected via direct imaging. However, neither GI nor CA is thought to operate in such a way that allows hot Jupiters to form in situ at their current locations close to the host stars. The disk cannot fragment at those distances, and cores with sufficient mass to attract significant envelopes cannot form. Therefore, the existence of hot Jupiters requires some form of "migration" from their original formation locations to their present orbits (see Section 2.3.2).

Migration may occur relatively early in the planet's history via the planet's interaction with, and transport through, the protoplanetary disk while the gas in the disk is still present (Lin et al., 1996). Alternately, migration may also occur at any time via scattering (Rasio and Ford, 1996) or secular interactions, such as Kozai resonances (Fabrycky and Tremaine, 2007), of the planet with other massive planetary or stellar components in the system. Measurements of orbital obliquities, i.e. the degree of alignment between the stellar equatorial plane and the planetary orbital plane, have been proposed to distinguish between the two migration scenarios (Gaudi and Winn, 2007). Whereas migration of a planet through a viscous disk would be expected to damp any initial misalignment, migration by scattering or Kozai resonances could lead to very high spin-orbit misalignments. The observations of a significant number of large spin-orbit misalignments in hot Jupiter systems in recent years initially supported the role of migration by scattering phenomena (Winn et al., 2010; Triaud et al. 2010). However, more recent studies have shown that spin-orbit misalignments can also be caused by planet migration through disks, which are themselves, misaligned due to torques induced by a distant stellar companion (Crida and Batygin, 2014). Consequently, dynamical measurements alone have not been able to conclusively constrain the formation of hot Jupiters, though directly imaged planets at wide separations (≳10 AU) seem more likely to be formed via GI.

*Chemical constraints*
Atmospheric elemental abundances of solar-system giant planets have led to important constraints on the origins of the solar system. For example, the observed super-solar enrichments of C, S, N, and the heavy noble gases (Section 2.2.1) support the formation of Jupiter and Saturn by core-accretion (see Sections 2.1 and 2.2.1). However, the oxygen abundance, which is a critical parameter in formation models, is not known in Saturn and Jupiter (see Sections 2.2.1 and 2.6). On the other hand, as discussed in Section 2.5.2, given the high temperatures of currently known giant exoplanets (T ~1000-3000 K), several key molecules are expected to be observable in their atmospheres and allow estimations of elemental abundance ratios involving H, C, O, and N. Nominal constraints on atmospheric C/H, O/H, and C/O ratios have already been reported for a few exoplanets and reveal both oxygen-rich (C/O < 1) as well as carbon-rich (C/O ≥ 1) compositions; the solar composition is oxygen-rich with C/O = 0.5.

Findings of super-solar C/O ratios in giant exoplanets are beginning to motivate new ideas on their formation mechanisms. The C/O ratios of most planet-hosting stars in the solar neighborhood are solar-like, i.e. oxygen-rich (e.g. Delgado-Mena et al., 2010). Thus, in the standard core-accretion model of planet formation, it is expected that oxygen-rich planetesimals with abundant $H_2O$ ice would dominate the planetesimal composition. Thus, the possibility of C-rich giant planet atmospheres orbiting O-rich stars poses a challenge to standard formation models of Jupiter and Saturn. An early investigation into this question was pursued in the context of Jupiter in the solar-system for which, as discussed above, only a lower limit on the O/H is known, which may allow for the possibility of C/O > 1. Lodders (2004) suggested the possibility of Jupiter forming by accreting tar-dominated planetesimals instead of those dominant in water ice as expected in the solar system based on the composition of minor bodies in the solar system. Following the inference of C/O ≥ 1 in the hot Jupiter WASP-12b (Madhusudhan et al., 2011a), Öberg et al. (2011b) suggested that C/O ratios in giant exoplanetary envelopes depend on the formation location of the planets in the disk relative to the ice lines of major C and O bearing volatile species, such as $H_2O$, CO, and $CO_2$. The C/O ratio of the gas in the nebula approaches 1 outside the CO and $CO_2$ ice lines. By predominantly accreting such C-rich gas, more so than O-rich planetesimals, gas giants could host C-rich atmospheres even when orbiting O-rich stars. It may also be possible that inherent inhomogeneities in the C/O ratios of the disk itself may contribute to higher C/O ratios of the planets relative to the host stars (Madhusudhan et al., 2011b). Additionally, the composition of the planet is also influenced by the temporal evolution of the chemical and thermodynamic properties of the disk at the formation location of the planet Saturn (Ali-Dib et al., 2014; Helling et al., 2014; Marboeuf et al., 2014). More recently, Madhusudhan et al. (2014c) suggested that O and C abundances of hot Jupiters could also provide constraints on their migration mechanisms. In particular, hot Jupiters with sub-solar elemental abundances are more likely to have migrated to their close-in orbits by disk-free mechanisms (e.g. scattering) rather than through the disk, regardless of their formation by core accretion or gravitational instability process.

Thus, various scenarios of giant planet formation and migration predict different limits on the metallicites and C/O ratios of giant exoplanets, which are testable with future high-precision and high-resolution observations of their atmospheres as will be possible with facilities like the *James Webb Space Telescope*, large ground-based telescopes of the future and dedicated space missions. As tighter constraints on the elemental abundances in exoplanets become available, investigating them together with elemental abundances in Saturn and Jupiter will allow development of convincing scenarios of the formation of gas giant planets in the solar system and extrasolar systems.

## 2.6 Outstanding issues and looking to the future

Existing observations of Saturn, its atmosphere, rings, and the moons have provided tantalizing clues into the formation and evolution scenarios of the Saturnian system. Additional insight has come from volatile composition and abundance data of giant exoplanets. Yet, the current observational constraints for developing robust models are either inadequate, poor, or simply non-existent, including those needed to address such fundamental questions as "does Saturn have a core today", "how does the size of Saturn's core compare to Jupiter's core", "what's Saturn's true intrinsic rotation rate", "what's Saturn's bulk composition, in particular the abundance of heavy elements, and how does it compare with Jupiter's bulk composition", "what's the helium abundance in the troposphere of Saturn", "is the history of heavy noble gases different from other heavy elements", and "what are the isotope ratios of H, He, N, S, O, Ar, Kr and Xe and what are their implications". New types of observations are required to address these issues. In the near future, the Cassini Grand Finale Mission appears promising for answering some of these questions.

Following a spectacular tour of the Saturnian system since reaching Saturn in 2004, the Cassini orbiter will enter its final phase of the mission in 2016, aptly named The Cassini Grand Finale, before the spacecraft crashes and burns in Saturn's atmosphere mid-2017. In the final 22 proximal orbits, Cassini's trajectory will take it high above the north pole, flying outside the F- ring and then plunging between Saturn and its innermost ring, skimming as close as ~1700 km above Saturn's cloud tops. These proximal orbits will give an unprecedented opportunity to carry out high precision measurements of higher order moments of gravity and magnetic fields, and the ring mass and particle distribution. These observations will provide useful constraints on the internal structure, rotation rate, and the age of Saturn's rings. As the orbits of the Juno spacecraft at Jupiter will be very similar to Cassini proximal orbits, a comparison between Jupiter and Saturn results in terms of the gravitational and magnetic fields will be possible. This extraordinary opportunity to gather comparable data on Jupiter and Saturn will not only help us to understand the intrinsic differences between these bodies but also get a sense of the variation we might expect among extrasolar giant planets within the same stellar system. The atmospheric composition relevant to Saturn's formation models requires in situ measurements, however.

Bulk composition and the atmospheric isotope determination of the giant planets cannot be carried out by remote sensing for the most part. The abundances of He and the heavy elements C, N, S, O, Ne, Ar, Kr, Xe, and isotope ratios D/H, $^3$He/$^4$He, $^{13}$C/$^{12}$C, $^{15}$N/$^{14}$N, $^{34}$S/$^{32}$S, $^{18}$O/$^{16}$O and the isotope ratios of the heavy noble gases are crucial constraints on the formation models. With the exception of carbon, their determination requires an entry probe at Saturn, as was done at Jupiter with the Galileo probe in 1995 [note the remote sensing result on D/H at Saturn is imprecise]. A shallow entry probe to 10 bars at Saturn is expected to deliver meaningful data on all of the above elements and isotopes, except perhaps oxygen, unless O/H is substantially sub-solar in Saturn. This is evident from Figure 2.9 showing the equilibrium cloud condensation levels of the condensible volatiles in Saturn's troposphere. For solar O/H, the base of the water cloud is found to be at 10 bars (cloud densities in the figure are *upper limits*; cloud bases are robust, however). As discussed earlier, water may be enriched similar to carbon, i.e. roughly 10× solar. In that case the base of the water cloud would be at ~20 bars. Because of convective and dynamical processes, well-mixed water may not be reached above two to three times these pressure levels, however. Thus, even in the unlikely scenario of solar water, probe measurements to at least 20-30 bars only can ensure reliable O/H determination in Saturn. If water is ten times solar, measurements down to at least 50 bars, preferably 100 bars, will be required for the O/H determination. If water in Saturn is greatly sub-solar, probes to ten bars will be able to determine the O/H directly in Saturn. Deep probes to such extreme environments of high pressures and temperatures and large radio opacity are presently unfeasible. However, Juno-like microwave radiometry from orbit at Saturn has the potential of determining the O/H and map deep water (and ammonia) abundance over the planet. Although the O/H determination in Saturn is desirable, its absence due to technical hurdles or cost constraints would not be a disaster. Comparison of all other elements and isotopes in Saturn, particularly the noble gases, with those in Jupiter measured by the Galileo probe and Juno's O/H would establish a trend or pattern from one gas giant planet to the other that may still provide meaningful constraints on Saturn's O/H. Other reservoirs of oxygen, such as CO, though much less abundant than $H_2O$, may also be exploited to obtain clues to the limits of O/H in Saturn. Future ground-based microwave measurements with improved capability are also promising for the deep water abundance. Refer to the chapter by Baines et al. for details on future exploration of Saturn.

Finally, composition data including especially the profiles of $H_2O$, CO and $CH_4$ in the atmospheres of giant exoplanets can provide a useful guide for Saturn. Similarly, in many respects, Saturn and Jupiter are ideal analogs for similar sized exoplanets around sun-like stars, despite the differences in their current orbital distances and resulting temperatures. Spectroscopic characterization of exoplanet atmospheres is proceeding rapidly, and there is a good prospect of addressing many of the outstanding issues including temperature structure and aerosol distribution. A comparison between atmospheric properties of a multitude of giant exoplanets is also essential. This chapter demonstrates that cross-fertilization between the giant planet research and the giant exoplanet research is beneficial both fields, and leads to a deeper understanding of the origin and evolution of this solar system and the extrasolar systems.

**Acknowledgments:** Discussions with several colleagues on the Juno, Cassini and Galileo science teams and ground-based planetary astronomers were beneficial in preparing this chapter. Joong Hyun In, Gloria Kim and Carmen Lee assisted with references and formatting.